\documentclass[final, oneside]{IEEEtran}

\usepackage[margin=2.2cm, centering]{geometry}
\usepackage{multirow}
\usepackage{booktabs}
\usepackage{lipsum}
\usepackage{amsfonts}
\usepackage{placeins}
\usepackage{array}
\usepackage{amsmath,cases,amssymb}
\usepackage{graphicx}
\usepackage{cite}
\usepackage{subfigure}
\usepackage{epsfig}
\usepackage{url}
\usepackage{algorithm}
\usepackage{algorithmic}
\usepackage{epstopdf}
\usepackage{balance}
\usepackage{color}

\newcommand{\ds}{\displaystyle}

\newtheorem{remark}{Remark}
\newcommand{\la}{\langle}
\newcommand{\ra}{\rangle}

\newcommand{\bw}{\mathbf{w}}
\newcommand{\bh}{\mathbf{h}}

\newcommand{\st}{{\mathrm{s.t.}}}

\newcommand{\clL}{{\cal L}}

\newcommand{\clI}{{\cal I}}
\newcommand{\clJ}{{\cal J}}

\newcommand{\clR}{{\cal R}}

\newcommand{\clS}{{\cal S}}

\newcommand{\Col}{{\sf Col}}

\newcommand{\bx}{\mathbf{x}}
\newcommand{\bfy}{\mathbf{y}}

\newcommand{\inter}{{\sf IN}}
\newcommand{\xk}{x^{(\kappa)}}
\newcommand{\yk}{y^{(\kappa)}}
\newcommand{\bxk}{\bx^{(\kappa)}}
\newcommand{\byk}{\bfy^{(\kappa)}}
\newcommand{\bwk}{\bw^{(\kappa)}}
\newcommand{\bwIk}{\bw^{I,(\kappa)}}

\newcommand{\balpha}{\pmb{\alpha}}

\ifCLASSOPTIONpeerreview

\fi
\begin{document}

\setcounter{page}{1}

\title{NOMA for throughput and EE maximization in Energy Harvesting Enabled Networks}
\author{A. A. Nasir, H. D. Tuan, T. Q. Duong, and M. Debbah
\thanks{Ali A. Nasir is with the Department of Electrical Engineering, King Fahd University of Petroleum and Minerals (KFUPM), Dhahran, Saudi Arabia (email: anasir@kfupm.edu.sa)}
\thanks{Hoang D. Tuan is with the School of Electrical and Data Engineering,
University of Technology, Sydney, NSW 2007, Australia (email:
tuan.hoang@uts.edu.au).}
\thanks{Trung Q. Duong is with the School of Electronics, Electrical Engineering and Computer Science, Queen's University Belfast, Belfast BT7 1NN, United Kingdom (e-mail: trung.q.duong@qub.ac.uk).}
\thanks{Merouane Debbah is with the Mathematical and Algorithmic Sciences Laboratory, France Research Center, Huawei Technologies, France (email: merouane.debbah@huawei.com).}
}
\maketitle
\ifCLASSOPTIONpeerreview
\vspace*{-1.cm}
\else
\fi
\begin{abstract}
Wireless power transfer via radio-frequency (RF) radiation is regarded as a potential solution to energize energy-constrained users, who are deployed close to the base stations (near-by users). However, energy transfer requires much more transmit power than normal information transfer, which makes it very challenging to provide the quality of service in terms of throughput for all near-by users and cell-edge users. Thus, it is of practical interest to employ non-orthogonal multiple access (NOMA) to improve the throughput of all network users, while fulfilling the energy harvesting requirements of the near-by users. To realize both energy harvesting and information decoding, we consider a transmit time-switching (transmit-TS) protocol. We formulate two important beamfoming problems of users' max-min throughput optimization and energy efficiency maximization under power constraint and energy harvesting thresholds at the nearly-located users. For these problems, the optimization objective and energy harvesting are non-convex in beamforming vectors. Thus, we develop efficient path-following algorithms to solve them. In addition, we also consider conventional power splitting (PS)-based energy harvesting receiver. Our numerical results confirm that the proposed transmit-TS based algorithms clearly outperform PS-based algorithms in terms of both, throughput and energy efficiency.
\end{abstract}

\section{Introduction}\label{sec:intro}

\subsection{Motivation}

To address the energy crisis of the enormous number of wirelessly connected devices, radio frequency (RF) energy harvesting/scavenging has emerged as a potential technology, which converts the energy of the received RF signals into electricity. This can be visualized as an opportunity for wireless communication system to not only deliver information but also energy to the near-by users, the ones that are located close to the base station (BS) \cite{Lu-14-A,Nasir-16-CL-A,Nguyen-17-May-A}. By means of the current state-of-the-art electronics, the received signal cannot be used for energy harvesting and  information decoding simultaneously. Thus, to realize both wireless energy harvesting (EH) and information decoding (ID), the user's receivers need to split the received signal for EH and ID either by power splitting (PS) or time switching (TS), where the latter one can be refered to as ``receive-TS" \cite{Lu-14-A,Tam-17-May-A,Nasir-16-TSP-A}. Though the PS approach has been shown to mostly outperform the receive-TS approach, however, the PS is complicated and inefficient for practical implementation. Recent findings in \cite{Tam-17-May-A,Nasir-16-TCOM-A} and \cite{Nasir-17-Nov-A} demonstrate the advantages of new ``transmit-TS" approach over PS approach, where information and energy is transferred separately and energy-constrained users' receivers do not need any sophisticated device.

During the information transmission, one of the most critical tasks is to provide quality of service (QoS) in terms of throughput to the
users, who are located far from the BS. Non-orthogonal multiple access (NOMA) technique (see e.g. \cite{Saietal13} and \cite{Shin-17-Oct-A}) is able to improve the throughput at far users by allowing the near-by users to access the information intended for the
far users. An efficient beamforming design for NOMA multicell systems has been recently proposed \cite{Nguyen-17-Dec-JSAC-A}.

\subsection{Literature Survey}

Very recently, some studies have been made regarding energy harvesting enabled NOMA systems. In \cite{Diamantoulakis-16-Dec-A} and \cite{Chingoska-16-Dec-A}, the authors considered a wireless powered communication network, where a BS charges users and enables them to transmit information during uplink communication by using NOMA scheme. Similar study was done by the authors in  \cite{Song-18-EA-A} for wireless-powered sensor network by using a NOMA scheme. In \cite{Moltafet-A-EA-18}, the authors considered a NOMA based heterogeneous network and studied the trade-off among the energy efficiency, fairness, harvested energy, and system sum rate. In \cite{Wang-A-EA-17}, the authors designed optimal resource allocation strategies for cognitive radio networks with NOMA. The authors considered non-linear EH model, where secondary users either harvest energy or decode information. The authors in \cite{Yang-17-INFOCOM-P} and \cite{Yang-17-EA-A} considered wireless power transfer and NOMA in machine-to-machine communication, where machine-type communication device harvests energy in the downlink while transmits information to BS via machine-type communication gateway in the uplink. {\color{black}In all the above works \cite{Diamantoulakis-16-Dec-A,Chingoska-16-Dec-A,Song-18-EA-A,Moltafet-A-EA-18,Wang-A-EA-17,Yang-17-INFOCOM-P,Yang-17-EA-A}, the authors assume single antenna nodes and consider a simple power allocation problem. Energy harvesting is more practical with multiple antenna beamforming, however, the above works \cite{Diamantoulakis-16-Dec-A,Chingoska-16-Dec-A,Song-18-EA-A,Moltafet-A-EA-18,Wang-A-EA-17,Yang-17-INFOCOM-P,Yang-17-EA-A} do not consider the practical and complex problem of multi-antenna beamforming. In addition, wireless nodes have to either harvest energy or decode information \cite{Diamantoulakis-16-Dec-A,Chingoska-16-Dec-A,Song-18-EA-A,Wang-A-EA-17,Yang-17-INFOCOM-P,Yang-17-EA-A}, and do not need to implement both EH and ID.}

In energy harvesting based cooperative NOMA systems, the cell-centered or ``nearly-located users" harvest energy from the wireless signals received from the BSs and act as a relay to forward the information to the ``far-located users" \cite{Ashraf-17-EA-A,Han-16-IET-A,Ha2017,Liu-16-Apr-A,Xuetal17,Yang-17-Jul-A,Sun-16-IET-A,Zhang-17-Dec-A,Do-17-A-EA,info8030111,Alsaba-18-EA-A}. Mostly, the authors considered PS approach at the relay users \cite{Ashraf-17-EA-A,Han-16-IET-A,Ha2017,Liu-16-Apr-A,Xuetal17,Yang-17-Jul-A,Sun-16-IET-A,Zhang-17-Dec-A,Alsaba-18-EA-A} and showed different analyses, e.g., deriving outage expressions for far end users \cite{Ashraf-17-EA-A,Han-16-IET-A,Liu-16-Apr-A,Zhang-17-Dec-A,Yang-17-Jul-A}, proposing different selection schemes for users pairing in NOMA \cite{Liu-16-Apr-A}, optimizing receiver PS ratios and transmit beamforming vectors \cite{Xuetal17,Sun-16-IET-A,Alsaba-18-EA-A}, specifically with the objective of maximizing the data rate of the strong user while satisfying the QoS requirement of the weak user \cite{Xuetal17}, or studying antenna selection schemes at the base station (BS) \cite{Zhang-17-Dec-A}. Recently, the authors in \cite{Do-17-A-EA} also considered hybrid receive-TS/PS approach for energy harvesting at the relay users. However, it is not practically easy to implement variable range power splitter. Reference \cite{info8030111}  considered simple time-switching (TS) mode at the relay receiver, where half of the time is dedicated for energy harvesting.

Exploiting wireless energy harvesting  in NOMA systems, the authors in \cite{Gong-17-Dec-A} considered various energy harvesting protocols, e.g., conventional receive-TS, PS and generalized (hybrid TS/PS), however, they assumed a very simple setup of single antenna BS and single user pair and did not consider multi-antenna beamforming. In \cite{He-17-Oct-P}, the authors also studied wireless power transfer and information transmission in NOMA system, however, they also assumed a simple setup with single antenna BS and employed PS based strategy for energy harvesting. {\color{black}In \cite{Deng-18-Apr-A}, the authors considered an energy-constrained full-duplex information transmitter, which harvests energy from a dedicated energy transmitter while transmitting information to the information receivers by employing NOMA. In \cite{Zhou-18-EA-A}, the authors considered cognitive radio based wireless information and power transfer network with NOMA. However, these works only considered transmit power minimization problem which is easier to solve compared to the challenging energy efficiency maximization problem, as will be considered in this work.}

\subsection{Research Gap and Contribution}

{\color{black}Recently, the authors in \cite{Xuetal17} solved the throughput optimization problem with multi-antenna beamforming and PS based EH in simple NOMA systems with one BS serving one nearly-located user and one far-located user. They {\color{black}and also the authors in \cite{Deng-18-Apr-A} and \cite{Zhou-18-EA-A}} employed semidefinite relaxation (SDR) to solve their respective beamforming-vector optimization problem, which is computationally inefficient as SDR has to first solve for matrix optimization \cite{Nasir-16-TCOM-A,Phan-12-A}. In addition, the authors in \cite{Xuetal17} employed a PS-based approach for energy harvesting, which is also not very practical due to the need of variable power splitting. Thus, an efficient solution for such important throughput maximin optimization problem is still missing. Meanwhile, energy efficiency (EE), which is defined as the sum throughput per Joule of consumed energy (ratio of sum throughput to power consumption), is an important key performance indicator for new wireless technologies \cite{Caval14}. Especially, energy harvesting brings in conflicting requirements from the viewpoint of EE, as it requires a stronger transmit power. Thus, energy efficiency maximization is an important research problem in EH enabled NOMA systems. To the best of authors' knowledge, the important problem of energy efficiency maximization with multi-antenna beamforming in EH enabled NOMA network has not been addressed in the literature. It is more difficult to solve the EE maximization problem than the throughput maximization problem due to the additional optimization variables appearing in the denominator of the EE function.

In this article, we consider multi-antenna beamforming in energy harvesting enabled NOMA systems. To achieve wireless energy harvesting, we do not prefer PS-based approach because it requires energy to be harvested from information-bearing signal, which needs a sophisticated energy harvesting device with impractical variable power splitter. Keeping in view the need to keep the receiver simple, we propose to consider a ``transmit-TS" approach, where information and energy are transmitted separately in fractional times, enabling the energy harvesting by simple devices. Thus, both throughput and harvested energy can be improved by separately designed information and energy beamformers. We formulate two important problems of worst-user throughput maximization and energy efficiency maximization under power constraint and energy harvesting constraints at the nearly-located users. For these problems, the optimization objective and energy harvesting constraints are highly non-convex, thus, we develop efficient path-following algorithms to solve them. For comparison purpose, we consider both PS-based and transmit-TS based energy harvesting and propose novel algorithms to solve above-mentioned novel beamforming problems.} Our numerical results confirm that the proposed transmit-TS approach clearly outperforms the PS approach in terms of both, throughput and energy efficiency.
\subsection{Organization and Notation}

The paper is organized as follows. After a brief system model, Section II presents the formulation of throughput and EE maximization problems and their computational solution for PS-based NOMA implementation. Section III describes the formulation and solution of such problems  for transmit-TS based NOMA implementation. Section IV evaluates the performance of our proposed algorithms by numerical examples. Finally, Section V concludes the paper.

\textit{Notation.} Bold-faced lower-case letters, e.g., $\mathbf{x}$,
 are used for vectors and lower-case letters, e.g., $x$, are used for for scalars. $\mathbf{x}^{H}$, $\mathbf{x}^{T}$, and $\mathbf{x}^{*}$ denote Hermitian transpose, normal transpose,  and conjugate of the vector $\mathbf{x}$, respectively. $\|\cdot\|$  stands for the vector's Euclidean norm. $\mathbb{C}$ is the set of all complex numbers, and $\emptyset$ is an empty set. $\Re\{x\}$ denotes the real part of a complex number $x$. $\nabla f(\mathbf{x})$ is the gradient of function  $f(\cdot)$ at $\mathbf{x}$. Also, we define $\la \mathbf{x},\mathbf{y} \ra \triangleq \mathbf{x}^H \mathbf{y}$. $\Col[a_i]_{i\in \clI}$ arranges $a_i$, $i\in \clI$ in
 row-block. For instance
 \[
 \Col[a_i]_{i\in\{1,2\}}=\begin{bmatrix}a_1\cr
  a_2\end{bmatrix}.
 \]

\ifCLASSOPTIONpeerreview
 \begin{figure}[t]
    \centering
    \includegraphics[width=0.65 \textwidth]{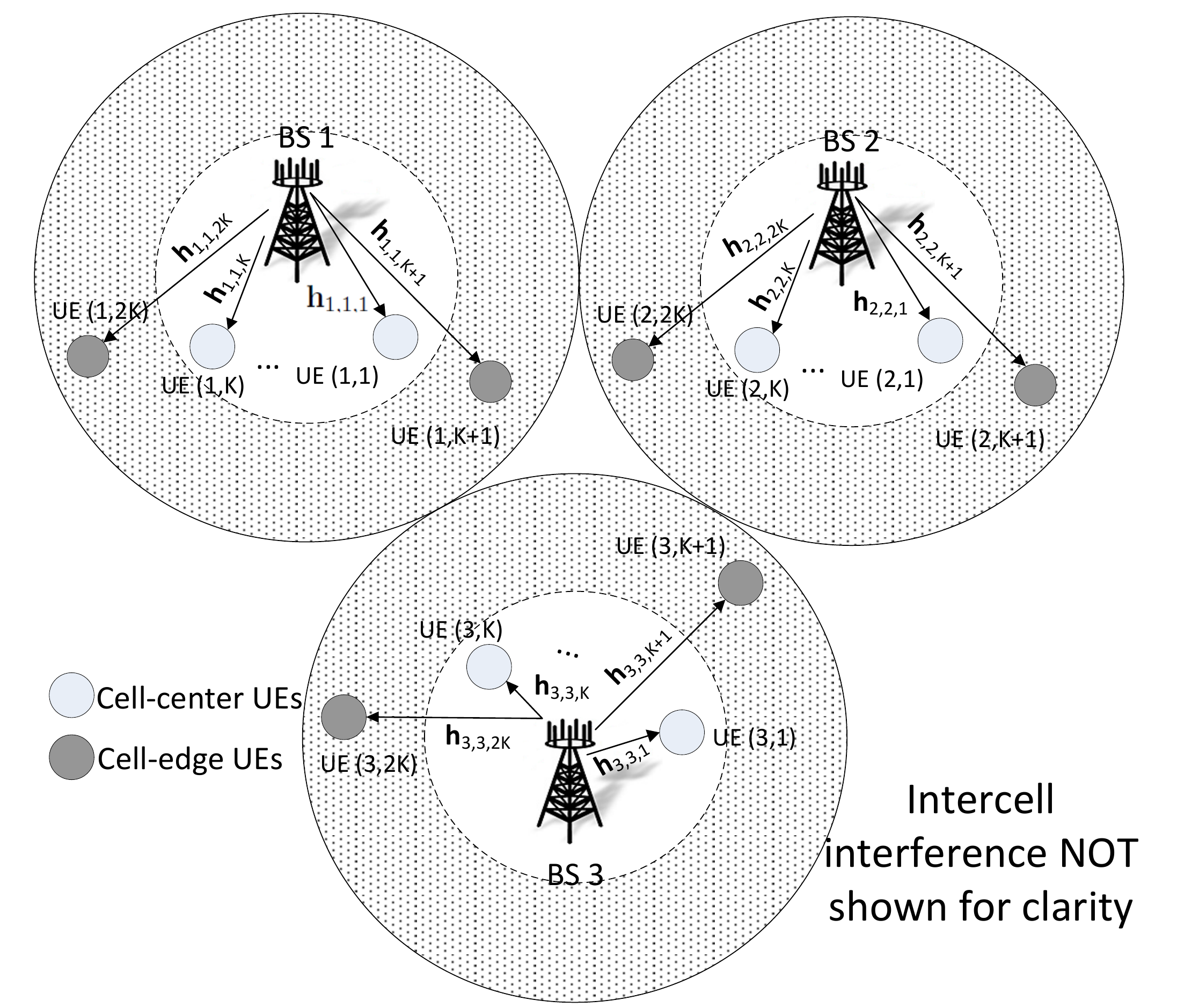} 
  \caption{{\color{black}Downlink multiuser multicell interference scenario in a dense network consisting of $K$ small cells. For clarity, the intercell interference channels are not shown, however, the interference occurs in all $K$ cells.}}
    \label{fig:sys_mod}
\end{figure}
\else
 \begin{figure}[t]
    \centering
    \includegraphics[width=0.49 \textwidth]{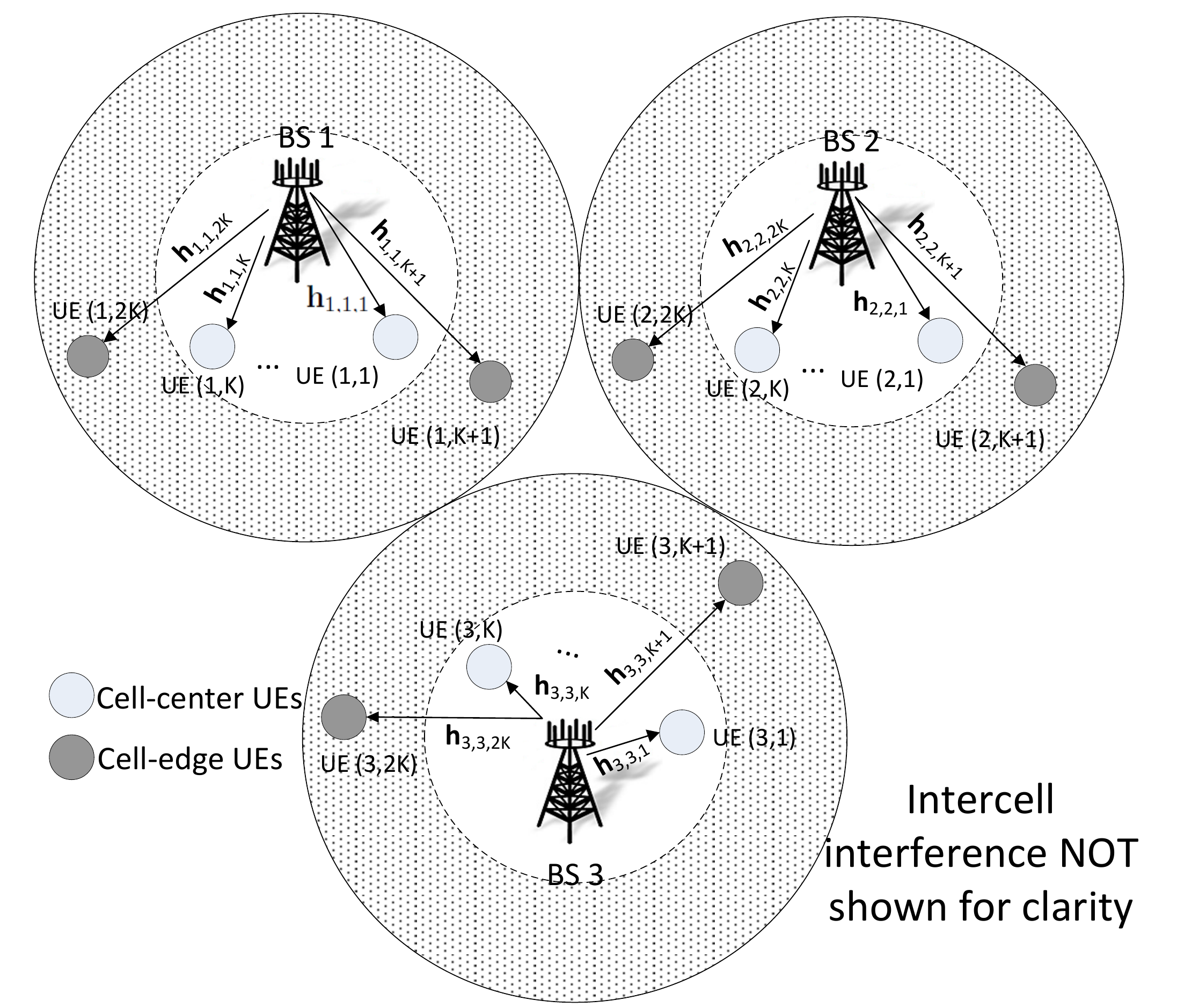} 
  \caption{{\color{black}Downlink multiuser multicell interference scenario in a dense network consisting of $K$ small cells. For clarity, the intercell interference channels are not shown, however, the interference occurs in all $K$ cells.}}
    \label{fig:sys_mod}
\end{figure}
\fi

\section{System Model and PS-based NOMA}\label{sec:sys_model}

Consider a downlink system consisting of $N$ cells under dense deployment, where the BS of each cell is equipped with $N_t$ antennas to serve $2K$ single-antenna-equipped users (UEs) within its cell. In each cell, there are $K$ near UEs (cell-center UEs), which are located inside the inner circular area and $K$ far UEs (cell-edge UEs), which are located in the ring area between inner circle and outer radius. A representative figure for $3$ cells is shown in Fig. \ref{fig:sys_mod}. The $K$ far UEs in each cell are not only in poorer channel conditions than other $K$ near UEs but also are under more drastic inter-cell interference from adjacent cells.

Upon denoting $\mathcal{I} \triangleq \{1, 2, \cdots, N\}$ and $\mathcal{J}  \triangleq  \{1, 2, \cdots, 2K\}$,
the $j$-th UE in the $i$-th cell is referred to as UE $(i,j)\in \mathcal{S}\triangleq \mathcal{I}\times\mathcal{J}$.
The cell-center UEs are UE $(i,j)$, $j\in\mathcal{J}_c\triangleq\{1,\cdots, K\}$ while the cell-edge UEs are
UE $(i,{\color{black}j})$, $j\in\mathcal{J}_e\triangleq\{K+1,\cdots, 2K\}$. Thus the set of cell-center UEs and the set of cell-edge UEs
are $\mathcal{S}_c\triangleq \mathcal{I}\times \mathcal{J}_c$ and $\mathcal{S}_e\triangleq \mathcal{I}\times \mathcal{J}_e$,
respectively.
Due their proximity, the cell-center UE $(i,j)$, $j\in\mathcal{J}_c$ is able to do both information decoding and energy
harvesting. Due the differentiated channel conditions between the cell-center and cell-edge UEs,  each
cell-center UE $(i,j)\in\mathcal{S}_c$ \textcolor{black}{is randomly paired} with cell-edge UE $(i,p(j))\in \mathcal{S}_e$ of the same cell to create
a virtual cluster to improve the network throughput. \footnote{\textcolor{black}{Using more sophisticated user-pairing strategies may improve the performance of   MIMO-NOMA networks (see e.g. \cite{DingTVT15}) but this is beyond the scope of this paper.}}
\subsection{PS-based NOMA}
For comparison point-of-view with transmit-TS based NOMA (will be presented in Section III), let us first develop the system model, problem formulation, and solution approach for PS-based NOMA implementation.

 The signal superpositions are precoded at the BSs prior to being transmitted to the UEs. Specifically, the message intended for UE $(i,j)$ is
$ s_{i,j} \in \mathbb{C}$ with $\mathbb{E}\{| s_{i,j}|^2\}=1$, which
is beamformed by vector $\mathbf{w}_{i,j}\in \mathbb{C}^{N_t}$. The received signals at UE $(i,j)$ and UE $(i,p(j))$ are expressed as
\begin{IEEEeqnarray}{rCl}
     y_{i,j} &=& \ds\sum_{(s,\ell)\in {\cal S}} \mathbf{h}_{s,i,j}\mathbf{w}_{s,\ell}  s_{s,\ell} + n_{i,j},\label{eq:yij}
 \end{IEEEeqnarray}
and
\begin{IEEEeqnarray}{rCl}
         y_{i,p(j)} &=& \ds\sum_{(s,\ell)\in {\cal S}} \mathbf{h}_{s,i,p(j)}\mathbf{w}_{s,\ell}  s_{s,\ell} + n_{i,p(j)},\label{eq:yipj}
\end{IEEEeqnarray}
where $\mathbf{h}_{s,i,j} \in\mathbb{C}^{1 \times N_t}$ is the MISO channel from the BS $s\in\mathcal{I}$ to UE $(i,j)\in\mathcal{S}$ and   $n_{i,j}\sim {\cal CN}(0,\sigma^2)$ is the additive noise. {\color{black}In this paper, we assume that full channel state information is available by some means, e.g., through coordination among BSs \cite{EmilBook}.}

To implement simultaneous wireless information and power transfer (SWIPT), the power splitter divides the received signal $y_{i,j}$ into two parts in the proportion of $\alpha_{i,j}:(1-\alpha_{i,j})$, where $\alpha_{i,j} \in (0,1)$ is termed as the PS ratio for UE $(i,j)$. The first part $\sqrt{\alpha_{i,j}} y_{i,j}$ forms an input to the ID receiver as:
\ifCLASSOPTIONpeerreview
\begin{equation}\label{swipt1}
 \sqrt{\alpha_{i,j}} y_{i,j} + z_{i,j}^c = \sqrt{\alpha_{i,j}}\left( \ds\sum_{(s,\ell)\in {\cal S}} \mathbf{h}_{s,i,j}\mathbf{w}_{s,\ell}  s_{s,\ell} + n_{i,j} \right)  + z_{i,j}^c,
\end{equation}
\else
\begin{eqnarray}\label{swipt1}
 \sqrt{\alpha_{i,j}} y_{i,j} + z_{i,j}^c &=& \notag \\
\sqrt{\alpha_{i,j}}\left( \sum_{(s,\ell)\in {\cal S}} \mathbf{h}_{s,i,j}\mathbf{w}_{s,\ell}  s_{s,\ell} + n_{i,j} \right)  + z_{i,j}^c,&&
\end{eqnarray}
\fi
where $z_{i,j}^c \sim  \mathcal{CN}(0,\sigma_c^2)$ is the additional noise introduced by the ID receiver circuitry.

The energy of the second part $\sqrt{1-\alpha_{i,j}} y_{i,j}$ of the received signal $y_{i,j}$ is harvested by the EH receiver of UE $(i,j)$ as
\begin{equation}\label{swipt2}
E_{i,j}(\mathbf{w},\alpha_{i,j})\triangleq \zeta_{i,j}(1-\alpha_{i,j})\left(p_{i,j}(\mathbf{w})+ \sigma^2\right),
\end{equation}
where we assume a linear EH model\footnote{The recently studied non-linear EH model and waveform design for
efficient wireless power transfer \cite{Clerckx-18-Feb-A} is beyond the scope of this work,
but could be incorporated in future research.} and for notational convenience, we define
$\mathbf{w} \triangleq  [\mathbf{w}_{i,j}]_{(i,j)\in\mathcal{S}}$, which constitutes all possible beamforming vectors. In \eqref{swipt2}, the constant $\zeta_{i,j} \in (0,1)$ denotes the efficiency of energy conversion at the EH receiver, and
\begin{align}\label{eq:pij}
p_{i,j}(\mathbf{w})=\ds\sum_{(s,\ell)\in {\cal S}}|\mathbf{h}_{s,i,j}\mathbf{w}_{s,\ell}|^2.
\end{align}
In NOMA, the message $s_{i,p(j)}$ is decoded by UE $(i,j)$ and $(i,p(j))$.
The interference plus noise at UE $(i,j)$ in decoding $s_{i,p(j)}$ is
\ifCLASSOPTIONpeerreview
\[
\begin{array}{lll}
\inter_{i,p(j)}^c(\mathbf{w})&\triangleq&\alpha_{i,j}\left(\ds\sum_{(s,\ell)\in{\cal S}\setminus\{(i, p(j))\}}
|\mathbf{h}_{s,i,j}\mathbf{w}_{s,\ell}|^2+ \sigma^2\right)+\sigma_c^2\nonumber\\
&=&\alpha_{i,j}(\|\clL_{i,p(j)}^c(\mathbf{w})\|^2+\sigma^2)+\sigma_c^2,\nonumber
\end{array}
\]
\else
\[
\begin{array}{rl}
\inter_{i,p(j)}^c(\mathbf{w})\triangleq&\alpha_{i,j}\left(\sum_{(s,\ell)\in{\cal S}\setminus\{(i, p(j))\}}
|\mathbf{h}_{s,i,j}\mathbf{w}_{s,\ell}|^2\right. \notag \\
&\left.+ \sigma^2\right) +\sigma_c^2\nonumber\\
=&\alpha_{i,j}(\|\clL_{i,p(j)}^c(\mathbf{w})\|^2+\sigma^2)+\sigma_c^2,\nonumber
\end{array}
\]
\fi
with
\begin{equation}\label{eqLMijb}
\clL_{i,p(j)}^c(\mathbf{w})\triangleq\Col[\mathbf{h}_{s,i,j}\mathbf{w}_{s,\ell}]_{(s,\ell)\in{\cal S}\setminus\{(i, p(j))\}},
\end{equation}
which is a linear operator. Therefore, the signal-to-interference-plus-noise (SINR) in decoding $s_{i,p(j)}$ at
UE $(i,j)$ is $|\mathbf{h}_{i,i,j}\mathbf{w}_{i,p(j)}|^2/\left(\|\clL_{i,p(j)}^c(\mathbf{w})\|^2+\sigma^2+\sigma_c^2/\alpha_{i,j}\right)
$.

Meanwhile, the edge-user UE $(i, p(j)$ decodes its own message $ s_{i,p(j)}$ only, so
the  interference plus noise at UE $(i,p(j))$ is
$\inter_{i,p(j)}^e(\mathbf{w})\triangleq\|\clL_{i,p(j)}^e(\mathbf{w})\|^2+ \sigma^2$
with the linear operator
\begin{equation}\label{eqLMipijb}
\clL_{i,p(j)}^e(\mathbf{w})\triangleq \Col[\mathbf{h}_{s,i,p(j)}\mathbf{w}_{s,\ell}]_{(s,\ell)\in{\cal S}\setminus \{(i, p(j))\}}.
\end{equation}
In what follows, we use the general rate function $\psi(\bx,\bfy,\nu)$ defined by
 \ifCLASSOPTIONpeerreview
\begin{equation}\label{basef}
\psi(\bx,\bfy,\nu)=\begin{cases}\begin{array}{ll}\ln \left(1+\|\bx\|^2/(\|\bfy\|^2+\sigma^2+\sigma_c^2/\nu ) \right)&\mbox{for}\quad \nu>0\cr \ln \left(1+\|\bx\|^2/(\|\bfy\|^2+\sigma^2) \right)&\mbox{for}\quad \nu=0.\end{array}\end{cases}
\end{equation}
\else
\begin{equation}\label{basef}
\psi(\bx,\bfy,\nu)=\begin{cases}\begin{array}{ll}\ln \left(1+ \frac{\|\bx\|^2}{\|\bfy\|^2+\sigma^2+\sigma_c^2/\nu} \right)&\mbox{for}\quad \nu>0\cr \ln \left(1+ \frac{\|\bx\|^2}{\|\bfy\|^2+\sigma^2} \right)&\mbox{for}\quad \nu=0.\end{array}\end{cases}
\end{equation}
\fi
{\color{black} Suppose that $\clR_{i,p(j)}$ is the achievable rate in decoding $ s_{i,p(j)}$. Since both near UE $(i,j)$ and far UE $(i,p(j))$ will decode the far user's information $ s_{i,p(j)}$, thus, $\clR_{i,p(j)}$ for UE $(i,j)$ and UE $(i,p(j))$, respectively, is given by}
\begin{eqnarray}
 \clR_{i,p(j)} &\leq&\psi(\mathbf{h}_{i,i,j}\bw_{i,p(j)},\clL^c_{i,p(j)}(\mathbf{w}),\alpha_{i,j}),\label{R1}\\
\clR_{i,p(j)}&\leq&\psi(\mathbf{h}_{i,i,j}\bw_{i,p(j)},\clL_{i,p(j)}^e(\bw),0).\label{R2}
 \end{eqnarray}
UE $(i,j)$ subtracts $s_{i,p(j)}$ from  the right hand side of (\ref{swipt1}) in decoding $s_{i,j}$. Then
the achievable rate $\clR_{i,j}$ by decoding $s_{i,j}$ is
 \begin{equation}
\clR_{i,j}\leq\psi(\mathbf{h}_{i,i,j}\bw_{i,j},\clL_{i,j}^c(\bw), \alpha_{i,j}),\label{R3}
 \end{equation}
with
\begin{equation}\label{lcij}
\clL_{i,j}^c(\mathbf{w})\triangleq\Col[\mathbf{h}_{s,i,j}\mathbf{w}_{s,\ell}]_{(s,\ell)\in{\cal S}\setminus\{(i, p(j)), (i,j)\}}.
\end{equation}
For convenience, we also use the notations
$
\balpha\triangleq (\alpha_{i,j})_{(i,j)\in\clS}$, $\balpha^{(\kappa)}\triangleq (\alpha^{(\kappa)}_{i,j})_{(i,j)\in\clS}$, $
\clR\triangleq (\clR_{i,j})_{(i,j)\in\clS}$, and
$\clR^{(\kappa)}\triangleq (\clR^{(\kappa)}_{i,j})_{(i,j)\in\clS}$.

We consider two basic problems:

$1)$ {\it Throughput maximin optimization}
\begin{subequations}\label{nnoma1}
\begin{eqnarray}
\ds\max_{\bw,\clR,\balpha}\ \min_{(i,j)\in\clI\times\clJ}\ \clR_{i,j}\quad\st\quad (\ref{R1}), (\ref{R2}), (\ref{R3}), \label{nnoma1a}\\
\ds\zeta_{i,j} \left(p_{i,j}(\mathbf{w}) + \sigma^2\right) \geq \frac{e^{\min}_{i,j}}{1-\alpha_{i,j}}, i\in\clI, j\in\clJ_c,\label{nnoma1b}\\
0<\alpha_{i,j}<1, i\in\clI, j\in\clJ_c,\label{nnoma1c}\\
\sum_{j\in \clJ} \|\mathbf{w}_{i,j}\|^2 \le P^{\max}_{i},\  i \in\mathcal{I}\label{nnoma1d}
\end{eqnarray}
\end{subequations}
where \eqref{nnoma1b} defines the EH constraint such that $e^{\min}_{i,j}$ is the EH threshold and $p_{i,j}(\mathbf{w})$ is defined in \eqref{eq:pij}, and $P^{\max}_i$ in \eqref{nnoma1d} is the transmit power budget of BS $i$.


$2)$ {\it Energy-efficiency maximization under QoS constraints}
 \ifCLASSOPTIONpeerreview
\begin{subequations}\label{nnoma2}
\begin{eqnarray}
\ds\max_{\mathbf{w},\clR,\balpha} \; \mathcal{F}(\mathbf{w},\clR)\triangleq\ds\frac{\sum_{(i,j)\in\clI\times\clJ}\ \clR_{i,j}}
{ \xi \pi(\bw)+P_c}\qquad \st\quad (\ref{R1}), (\ref{R2}), (\ref{R3}), (\ref{nnoma1b}), (\ref{nnoma1c}), (\ref{nnoma1d})\label{nnoma2a}\\
  \clR_{i,j} \geq r_{i,j},\   i \in\mathcal{I},\  j\in\mathcal{J}_c, \label{nnoma2b}\\
\clR_{i,p(j)} \geq r_{i,p(j)},\,   i \in\mathcal{I},\,  j\in\mathcal{J}_c,\label{nnoma2c}
 \end{eqnarray}
 \end{subequations}
 \else
 \begin{subequations}\label{nnoma2}
\begin{eqnarray}
\ds\max_{\mathbf{w},\clR,\balpha} \; \mathcal{F}(\mathbf{w},\clR)\triangleq\ds\frac{\sum_{(i,j)\in\clI\times\clJ}\ \clR_{i,j}}
{\xi \pi(\bw)+P_c} \notag \\
\qquad \st\quad (\ref{R1}), (\ref{R2}), (\ref{R3}), (\ref{nnoma1b}), (\ref{nnoma1c}), (\ref{nnoma1d})\label{nnoma2a}\\
  \clR_{i,j} \geq r_{i,j},\   i \in\mathcal{I},\  j\in\mathcal{J}_c, \label{nnoma2b}\\
\clR_{i,p(j)} \geq r_{i,p(j)},\ i \in\mathcal{I},\,  j\in\mathcal{J}_c,\label{nnoma2c}
 \end{eqnarray}
 \end{subequations}
\fi
where $\pi(\bw)=\sum_{i\in\clI}\sum_{j\in \clJ} \|\mathbf{w}_{i,j}\|^2$,  $\xi$ is the reciprocal constant power amplifier efficiency, $P_c \triangleq N_t P_A + P_\text{cir}$, $P_A$ is the power dissipation at each transmit antenna, $P_\text{cir}$ is the fixed circuit power consumption for base-band processing, and \eqref{nnoma2b} and \eqref{nnoma2c} are the quality-of-service constraints, such that, $r_{i,j}$, $ i \in\mathcal{I},\, \forall j\in\mathcal{J}$ is the threshold rate to ensure a certain quality-of-service.
\subsection{Computational solutions for PS-based NOMA}\label{sec:CQBI}
Let us first address the throughput maximin optimization problem \eqref{nnoma1}. We have to resolve the non-convex rate constraints (\ref{R1}), (\ref{R2}), and (\ref{R3}) and the non-convex EH constraint \eqref{nnoma1b}. In order to deal with the non-convexity of rate constraints (\ref{R1}), (\ref{R2}), and (\ref{R3}), we have to provide a concave lower bounding function for $\psi(\bx,\bfy,0)$ defined by (\ref{basef}), at a given point $(\bx^{(\kappa)}, \bfy^{(\kappa)})$ \cite{Nasir-16-TSP-A,Nasir-17-Nov-A}. In the Appendix A, we prove the following new and universal concave function bound:
\begin{equation}\label{es1}
\psi(\bx,\bfy,0)\geq \Lambda_0(\bx,\bfy)
\end{equation}
over the trust region
\begin{equation}\label{tr0}
2\Re\{(\bx^{(\kappa)})^H\bx\}-\|\bx^{(\kappa)}\|^2>0,
\end{equation}
with $\psi(\bx^{(\kappa)},\bfy^{(\kappa)},0)=\Lambda_0(\bx^{(\kappa)},\bfy^{(\kappa)})$,
where
\ifCLASSOPTIONpeerreview
\begin{eqnarray}
\Lambda_0(\bx,\bfy)
&\triangleq&a_0(\bx^{(\kappa)},\bfy^{(\kappa)})-\ds\frac{\|\bx^{(\kappa)}\|^2}{2\Re\{(\bx^{(\kappa)})^H\bx\}-\|\bx^{(\kappa)}\|^2}
\nonumber\\
&&-b_0(\bx^{(\kappa)},\bfy^{(\kappa)})\|\bx\|^2-c_0(\bx^{(\kappa)},\bfy^{(\kappa)})\|\bfy\|^2,\label{es1a}
\end{eqnarray}
\else
\begin{align}
\Lambda_0(\bx,\bfy)
&\triangleq a_0(\bx^{(\kappa)},\bfy^{(\kappa)})-\ds\frac{\|\bx^{(\kappa)}\|^2}{2\Re\{(\bx^{(\kappa)})^H\bx\}-\|\bx^{(\kappa)}\|^2}
\nonumber\\
&-b_0(\bx^{(\kappa)},\bfy^{(\kappa)})\|\bx\|^2-c_0(\bx^{(\kappa)},\bfy^{(\kappa)})\|\bfy\|^2,\label{es1a}
\end{align}
\fi
and
\ifCLASSOPTIONpeerreview
\begin{eqnarray}
a_0(\bx^{(\kappa)},\bfy^{(\kappa)})&=&\psi(\bx^{(\kappa)},\bfy^{(\kappa)})+2-\ds\frac{\|\bxk\|^2}{(\|\bxk\|^2+
\|\byk\|^2+\sigma^2)}\frac{\sigma^2}{\|\byk\|^2+\sigma^2} ,\label{es2}\\
0<b_0(\bx^{(\kappa)},\bfy^{(\kappa)})&=&\ds\frac{\|\byk\|^2+\sigma^2}{(\|\bxk\|^2+
\|\byk\|^2+\sigma^2)\|\bxk\|^2},\label{es3}\\
0<c_0(\bx^{(\kappa)},\bfy^{(\kappa)})&=&\ds\frac{\|\bxk\|^2}{(\|\bxk\|^2+
\|\byk\|^2+\sigma^2)(\|\byk\|^2+\sigma^2)} .\label{es4}
\end{eqnarray}
\else
\begin{align}
a_0(\bx^{(\kappa)},\bfy^{(\kappa)})&= \psi(\bx^{(\kappa)},\bfy^{(\kappa)})+2 \notag \\ &-\ds\frac{\|\bxk\|^2}{(\|\bxk\|^2+
\|\byk\|^2+\sigma^2)}\frac{\sigma^2}{\|\byk\|^2+\sigma^2} ,\label{es2}\\
b_0(\bx^{(\kappa)},\bfy^{(\kappa)}) &= \ds\frac{\|\byk\|^2+\sigma^2}{(\|\bxk\|^2+
\|\byk\|^2+\sigma^2)\|\bxk\|^2} \notag \\ & >0,\label{es3}\\
c_0(\bx^{(\kappa)},\bfy^{(\kappa)})
&= \ds\frac{\|\bxk\|^2}{(\|\bxk\|^2+
\|\byk\|^2+\sigma^2)(\|\byk\|^2+\sigma^2)} \notag \\ & >0 .\label{es4}
\end{align}
\fi
Analogously, we can derive the following concave lower bounding function for $\psi(\bx,\bfy,\mu)$  at a given point $(\bx^{(\kappa)}, \bfy^{(\kappa)}, \mu^{(\kappa)})$,
\begin{equation}\label{es.1}
\psi(\bx,\bfy,\mu)\geq \Lambda(\bx,\bfy,\mu)
\end{equation}
over the trust region (\ref{tr0}), with
$
\psi(\bx^{(\kappa)},\bfy^{(\kappa)},\mu^{(\kappa)})=\Lambda(\bx^{(\kappa)},\bfy^{(\kappa)},\mu^{(\kappa)})
$, where
\ifCLASSOPTIONpeerreview
\begin{eqnarray}
\Lambda(\bx,\bfy,\mu)
&\triangleq&a(\bx^{(\kappa)},\bfy^{(\kappa)},\mu^{(\kappa)})-
\ds\frac{\|\bx^{(\kappa)}\|^2}{2\Re\{(\bx^{(\kappa)})^H\bx\}-\|\bx^{(\kappa)}\|^2}
\nonumber\\
&&-b(\bx^{(\kappa)},\bfy^{(\kappa)},\mu^{(\kappa)})\|\bx\|^2-c(\bx^{(\kappa)},\bfy^{(\kappa)},\mu^{(\kappa)})
(\|\bfy\|^2+\sigma_c^2/\mu),\label{es.1a}
\end{eqnarray}
\else
\begin{align}
\Lambda(\bx,\bfy,\mu)&\triangleq a(\bx^{(\kappa)},\bfy^{(\kappa)},\mu^{(\kappa)}) \notag \\ &-
\ds\frac{\|\bx^{(\kappa)}\|^2}{2\Re\{(\bx^{(\kappa)})^H\bx\}-\|\bx^{(\kappa)}\|^2}
\nonumber\\
& -b(\bx^{(\kappa)},\bfy^{(\kappa)},\mu^{(\kappa)})\|\bx\|^2 \notag \\ & \hspace{0.5cm} -c(\bx^{(\kappa)},\bfy^{(\kappa)},\mu^{(\kappa)})
(\|\bfy\|^2+\sigma_c^2/\mu),\label{es.1a}
\end{align}
\fi
and
\ifCLASSOPTIONpeerreview
\begin{eqnarray}
a(\bx^{(\kappa)},\bfy^{(\kappa)},\mu^{(\kappa)})&=&\psi(\bx^{(\kappa)},\bfy^{(\kappa)},\mu^{(\kappa)})+2
-\ds\frac{\|\bxk\|^2}{\|\bxk\|^2+
\|\byk\|^2+\sigma^2+\sigma_c^2/\mu^{(\kappa)}}\nonumber\\
&&\ds\times\frac{\sigma^2}{\|\byk\|^2+\sigma^2+\sigma_c^2/\mu^{(\kappa)}},\label{es.2}\\
0<b(\bx^{(\kappa)},\bfy^{(\kappa)},\mu^{(\kappa)})&=&\ds\frac{\|\byk\|^2+\sigma^2+\sigma_c^2/\mu^{(\kappa)}}{(\|\bxk\|^2+
\|\byk\|^2+\sigma^2+\sigma_c^2/\mu^{(\kappa)})\|\bxk\|^2},\label{es.3}\\
0<c(\bx^{(\kappa)},\bfy^{(\kappa)},\mu^{(\kappa)})&=&\ds\frac{\|\bxk\|^2}{(\|\bxk\|^2+
\|\byk\|^2+\sigma^2+\sigma_c^2/\mu^{(\kappa)})(\|\byk\|^2+\sigma^2+\sigma_c^2/\mu^{(\kappa)})} .\label{es.4}
\end{eqnarray}
\else
\begin{align}
& \hspace{-0.5cm} a(\bx^{(\kappa)},\bfy^{(\kappa)},\mu^{(\kappa)}) = \psi(\bx^{(\kappa)},\bfy^{(\kappa)},\mu^{(\kappa)})+2
 \notag \\ &\hspace{2cm} - \ds\frac{\|\bxk\|^2}{\|\bxk\|^2+
\|\byk\|^2+\sigma^2+\sigma_c^2/\mu^{(\kappa)}}\nonumber\\
&\hspace{2cm} \ds\times\frac{\sigma^2}{\|\byk\|^2+\sigma^2+\sigma_c^2/\mu^{(\kappa)}},\label{es.2}\\
0<&b(\bx^{(\kappa)},\bfy^{(\kappa)},\mu^{(\kappa)}) \notag \\ =&\ds\frac{\|\byk\|^2+\sigma^2+\sigma_c^2/\mu^{(\kappa)}}{(\|\bxk\|^2+
\|\byk\|^2+\sigma^2+\sigma_c^2/\mu^{(\kappa)})\|\bxk\|^2},\label{es.3}\\
0<&c(\bx^{(\kappa)},\bfy^{(\kappa)},\mu^{(\kappa)}) =\notag \\  & \hspace{-0.6cm} \hspace{-0.15cm} \ds\frac{\|\bxk\|^2}{ \left(\|\bxk\|^2+
\|\byk\|^2+\sigma^2+ \frac{\sigma_c^2}{\mu^{(\kappa)}} \hspace{-0.1cm} \right) \hspace{-0.15cm} \left(\|\byk\|^2+\sigma^2+ \frac{\sigma_c^2}{\mu^{(\kappa)}} \hspace{-0.1cm} \right)} .\label{es.4}
\end{align}
\fi
Let $(\mathbf{w}^{(\kappa)}, \clR^{(\kappa)}, \balpha^{(\kappa)})$ be a feasible point for
(\ref{nnoma1}) that is found from the $(\kappa-1)$th iteration. Applying inequality (\ref{es.1}) yields
$\psi(\mathbf{h}_{i,i,j}\bw_{i,p(j)},\clL^c_{i,p(j)}(\mathbf{w}),\alpha_{i,j})\geq \Lambda(\mathbf{h}_{i,i,j}\bw_{i,p(j)},\clL^c_{i,p(j)}(\mathbf{w}),\alpha_{i,j})$ over the trust region
\ifCLASSOPTIONpeerreview
\begin{equation}
2\Re\{(\bwk_{i,p(j)})^H\mathbf{h}^H_{i,i,p(j)} \mathbf{h}_{i,i,p(j)}\bw_{i,p(j)}\}- \|\mathbf{h}_{i,i,p(j)}\bwk_{i,p(j)}\|^2
>0,\label{nnomatkf}
\end{equation}
\else
\begin{eqnarray}
2\Re\{(\bwk_{i,p(j)})^H\mathbf{h}^H_{i,i,p(j)} \mathbf{h}_{i,i,p(j)}\bw_{i,p(j)}\}&&\nonumber\\
- \|\mathbf{h}_{i,i,p(j)}\bwk_{i,p(j)}\|^2&>&0,\label{nnomatkf}
\end{eqnarray}
\fi
where
the concave function $\Lambda(\mathbf{h}_{i,i,j}\bw_{i,p(j)},\clL^c_{i,p(j)}(\mathbf{w}),\alpha_{i,j})$ is defined from (\ref{es.1a}), (\ref{es.2})-(\ref{es.4}) for $\bx^{(\kappa)}=\mathbf{h}_{i,i,j}\bw^{(\kappa)}_{i,p(j)}$,
$\bfy^{(\kappa)}=\clL^c_{i,p(j)}(\mathbf{w}^{(\kappa)})$, and $\mu^{(\kappa)}=\alpha_{i,j}^{(\kappa)}$. As such,
the nonconvex constraint (\ref{R1}) is innerly approximated  by the convex constraints (\ref{nnomatkf}) and
\begin{equation}
\clR_{i,p(j)}\leq\Lambda(\mathbf{h}_{i,i,j}\bw_{i,p(j)},\clL^c_{i,p(j)}(\mathbf{w}),\alpha_{i,j}).\label{nnomatkb}
\end{equation}
Analogously, the nonconvex constraints (\ref{R2}) and (\ref{R3}) are innerly approximated by the convex constraints
\ifCLASSOPTIONpeerreview
\begin{eqnarray}
\clR_{i,p(j)}\leq\Lambda_0(\mathbf{h}_{i,i,p(j)}\bw_{i,p(j)},\clL^e_{i,p(j)}(\bw)),\label{nnomatkc}\\
\clR_{i,j}\leq\Lambda(\mathbf{h}_{i,i,j}\bw_{i,j},\clL^c_{i,j}(\bw),\alpha_{i,j}),\label{nnomatkd}\\
2\Re\{(\bwk_{i,p(j)})^H\mathbf{h}^H_{i,i,j}\mathbf{h}_{i,i,j}\bw_{i,p(j)}\}-\|\mathbf{h}_{i,i,j}\bwk_{i,p(j)} \|^2
>0,\label{nnomatkg}\\
2\Re\{(\bwk_{i,j})^H\mathbf{h}^H_{i,i,j}\mathbf{h}_{i,i,j}\bw_{i,j}\}- \|\mathbf{h}_{i,i,j}\bwk_{i,j}\|^2>0,\label{nnomatkh}
\end{eqnarray}
\else
\begin{eqnarray}
\clR_{i,p(j)}\leq\Lambda_0(\mathbf{h}_{i,i,p(j)}\bw_{i,p(j)},\clL^e_{i,p(j)}(\bw)),\label{nnomatkc}\\
\clR_{i,j}\leq\Lambda(\mathbf{h}_{i,i,j}\bw_{i,j},\clL^c_{i,j}(\bw),\alpha_{i,j}),\label{nnomatkd}\\
2\Re\{(\bwk_{i,p(j)})^H\mathbf{h}^H_{i,i,j}\mathbf{h}_{i,i,j}\bw_{i,p(j)}\}\nonumber\\
-\|\mathbf{h}_{i,i,j}\bwk_{i,p(j)} \|^2>0,\label{nnomatkg}\\
2\Re\{(\bwk_{i,j})^H\mathbf{h}^H_{i,i,j}\mathbf{h}_{i,i,j}\bw_{i,j}\}- \|\mathbf{h}_{i,i,j}\bwk_{i,j}\|^2>0,\label{nnomatkh}
\end{eqnarray}
\fi
where the concave functions $\Lambda_0(\mathbf{h}_{i,i,p(j)}\bw_{i,p(j)},\clL^e_{i,p(j)}(\bw))$
 is defined from
(\ref{es1a})-(\ref{es4})
 for $\bx^{(\kappa)}=\mathbf{h}_{i,i,p(j)}\bw^{(\kappa)}_{i,p(j)}$,
$\bfy^{(\kappa)}=\clL^e_{i,p(j)}(\mathbf{w}^{(\kappa)})$,
while
the concave functions
$\Lambda(\mathbf{h}_{i,i,j}\bw_{i,j},\clL^c_{i,j}(\bw),\alpha_{i,j})$ is defined from
(\ref{es.1a})-(\ref{es.4})
 for $\bx^{(\kappa)}=\mathbf{h}_{i,i,j}\bw^{(\kappa)}_{i,j}$,
$\bfy^{(\kappa)}=\clL^c_{i,j}(\mathbf{w}^{(\kappa)})$, and $\mu^{(\kappa)}=\alpha_{i,j}^{(\kappa)}$.

Next, by using the following approximation \cite{Nasir-17-Nov-A}
\[
| \mathbf{h}_{s,i,j}^H \mathbf{x}|^2  \geq  -| \mathbf{h}_{s,i,j}^H \mathbf{x}|^2 + 2 \Re \left\{ \left(\mathbf{x}^{(\kappa)}\right)^H \mathbf{h}_{s,i,j} \mathbf{h}_{s,i,j}^H  \mathbf{x}\right\},
\]
an inner convex approximation for the nonconvex EH constraint \eqref{nnoma1b} is given by
\ifCLASSOPTIONpeerreview
\begin{equation}
\ds\zeta_{i,j}\sum_{(s,\ell)\in\clS}\left(2\Re\{(\bw_{s,\ell}^{(\kappa)})^H\bh^H_{s,i,j}\bh_{s,i,j}\bw_{s,\ell}  \}
-|\bh_{s,i,j}\bw^{(\kappa)}_{s,\ell}|^2 \right)+\zeta_{i,j}\sigma^2\geq \frac{e^{\min}_{i,j}}{1-\alpha_{i,j}}. \label{nnomatkba}
\end{equation}
\else
\begin{eqnarray}
\zeta_{i,j}\sum_{(s,\ell)\in\clS}\left(2\Re\{(\bw_{s,\ell}^{(\kappa)})^H\bh^H_{s,i,j}\bh_{s,i,j}\bw_{s,\ell}  \}\right.\nonumber\\
\left.-|\bh_{s,i,j}\bw^{(\kappa)}_{s,\ell}|^2 \right) +\zeta_{i,j}\sigma^2\geq \frac{e^{\min}_{i,j}}{1-\alpha_{i,j}}. \label{nnomatkba}
\end{eqnarray}
\fi
At the $\kappa$-th iteration, we solve the following convex quadratic optimization problem of computational
complexity $\mathcal{O} \left((2 K N (N_t + 2))^3  \left( 8 KN +  N \right) \right)$ \cite{Peaucelle-02-A}
to generate the next feasible point $(\bw^{(\kappa+1)},\clR^{(\kappa+1)},\balpha^{(\kappa+1)})$:
\begin{equation}\label{nnomatk}
\ds\max_{\bw,\balpha,\clR}\ \min_{(i,j)\in\clI\times\clJ}\ \clR_{i,j}\quad\st\quad (\ref{nnoma1c}), (\ref{nnoma1d}), (\ref{nnomatkf})-(\ref{nnomatkba}).
\end{equation}
Algorithm~\ref{alg3} outlines the steps to solve the throughput maximin optimization problem \eqref{nnoma1}. Note that $\min_{(i,j)\in\clI\times\clJ}\ \clR_{i,j}^{(\alpha+1)}>\min_{(i,j)\in\clI\times\clJ}\ \clR_{i,j}^{(\alpha)}$ because $(\bw^{(\kappa+1)},\clR^{(\kappa+1)},\balpha^{(\kappa+1)})$ is the optimal solution of (\ref{nnomatk}) while
 $(\bw^{(\kappa)},\clR^{(\kappa)},\balpha^{(\kappa)})$ is its feasible point. Therefore, Algorithm \ref{alg3} generates
 a sequence $\{ (\bw^{(\kappa)},\clR^{(\kappa)},\balpha^{(\kappa)})\}$ of improved feasible points for (\ref{nnoma1}).
 By using similar arguments to that shown in \cite{Nasir-16-TSP-A,Nasir-16-TCOM-A}, we can easily show that Algorithm~\ref{alg3} at least converges to a locally optimal solution of \eqref{nnoma1}, which satisfies the
 Karush-Kuhn-Tucker (KKT) optimality condition.

 Since the EH constraint (\ref{nnoma1b}) is nonconvex, locating a feasible point $(\mathbf{w}^{(0)}, \clR^{(0)}, \balpha^{(0)})$ for  (\ref{nnoma1}) is a nonconvex problem, which is resolved via the iterations
\ifCLASSOPTIONpeerreview
\begin{eqnarray}\label{qosA1}
\ds\max_{\mathbf{w},\balpha} \min_{(i,j)\in\clI\times\clJ_c}\ \bigg\{
\ds\zeta_{i,j}\sum_{(s,\ell)\in\clS}\left(2\Re\{(\bw_{s,\ell}^{(\kappa)})^H\bh^H_{s,i,j}\bh_{s,i,j}\bw_{s,l}  \}
-|\bh_{s,i,j}\bw^{(\kappa)}_{s,l}|^2 \right)
+\zeta_{i,j}\sigma^2-\frac{e^{\min}_{i,j}}{1-\alpha_{i,j}}  \bigg\}\quad
\nonumber\\
\st\quad \eqref{nnoma1c}, \eqref{nnoma1d}
\end{eqnarray}
\else
\begin{align}\label{qosA1}
\ds\max_{\mathbf{w},\balpha} \min_{(i,j)\in\clI\times\clJ_c} &  \bigg\{ \ds\zeta_{i,j}\sum_{(s,\ell)\in\clS}\left(2\Re\{(\bw_{s,\ell}^{(\kappa)})^H\bh^H_{s,i,j}\bh_{s,i,j}\bw_{s,l}  \} \right. \notag \\ & \left.
-|\bh_{s,i,j}\bw^{(\kappa)}_{s,l}|^2 \right)
 +\zeta_{i,j}\sigma^2-\frac{e^{\min}_{i,j}}{1-\alpha_{i,j}}  \bigg\} \notag \\
\st &\quad \eqref{nnoma1c}, \eqref{nnoma1d}
\end{align}
\fi
till reaching a value more than or equal to $0$ and thus the feasibility for (\ref{nnoma1b}). We
then reset   $(\mathbf{w}^{(0)}, \balpha^{(0)})$ to this feasible point and
define $\clR_{i, p(j)}^{(0)}$ as
the minimum among the values of the right-hand sides (RHS) of  (\ref{R1}) and (\ref{R2}) at $(\mathbf{w}^{(0)}, \balpha^{(0)})$, and $\clR_{i, j}^{(0)}$ as the value of the RHS of (\ref{R3}) at $(\mathbf{w}^{(0)}, \balpha^{(0)})$
to have a feasible point $(\bw^{(0)}, \clR^{(0)}, \balpha^{(0)})$ of \eqref{nnoma1} for initializing Algorithm
\ref{alg3}.
\begin{algorithm}[t]
\begin{algorithmic}[1]
\protect\caption{PS-based algorithm for throughput max-min optimization problem \eqref{nnoma1}}
\label{alg3}
\global\long\def\algorithmicrequire{\textbf{Initialization:}}
\REQUIRE  Set $\kappa:=0$ and initialize a feasible point $(\bw^{(0)}, \clR^{(0)}, \balpha^{(0)})$ for \eqref{nnoma1}
\STATE Until convergence of the objective in \eqref{nnoma1} repeat:
 Solve the convex  optimization problem (\ref{nnomatk}) to generate the feasible point
$(\mathbf{w}^{(\kappa+1)}, \clR^{(\kappa+1)}, \balpha^{(\kappa+1)})$ for (\ref{nnoma1}).
 Reset $\kappa:=\kappa+1$.
\end{algorithmic} \end{algorithm}

Next, we address the energy-efficiency maximization problem (\ref{nnoma2}), which is equivalent to the
following minimization problem
\ifCLASSOPTIONpeerreview
\begin{eqnarray}\label{nnoma2e}
\ds\min_{\bw, \balpha, \clR}\
\frac{\xi \pi(\bw)+P_c}{\sum_{(i,j)\in\clI\times\clJ}\ \clR_{i,j}}
\quad\st\quad
(\ref{R1})-(\ref{R3}), (\ref{nnoma1b})-(\ref{nnoma1d}), (\ref{nnoma2b}), (\ref{nnoma2c}).
\end{eqnarray}
\else
\begin{eqnarray}\label{nnoma2e}
\ds\min_{\bw, \balpha, \clR}\
\frac{\xi\pi(\bw)+P_c}{\sum_{(i,j)\in\clI\times\clJ}\ \clR_{i,j}}
\quad\st \nonumber\\
(\ref{R1}), (\ref{R2}), (\ref{R3}), (\ref{nnoma1b}), (\ref{nnoma1c}), (\ref{nnoma1d}), (\ref{nnoma2b}), (\ref{nnoma2c}).
\end{eqnarray}
\fi
The advantage of this transformation is that the objective function is already convex so there is no need of using
Dinkelbach's iteration or approximating it. Actually, the Dinkelbach iteration for (\ref{nnoma2}) still involves
a nonconvex optimization problem, which is as computationally difficult as (\ref{nnoma2}) itself (see e.g., \cite{Buzzi-16-Apr-A}). At the $\kappa$-th iteration,
 we solve the following convex optimization problem of computational complexity $\mathcal{O} \left((2 K N (N_t + 2))^3  \left( 10 KN +  N \right) \right)$
  to generate its next feasible point $(\bw^{(\kappa+1)}, \clR^{(\kappa+1)}, \balpha^{(\kappa+1)})$:
\ifCLASSOPTIONpeerreview
\begin{eqnarray}\label{nnomatk2}
\ds\min_{\bw, \balpha, \clR}\
\frac{ \xi \pi(\bw)+P_c}{\sum_{(i,j)\in\clI\times\clJ}\ \clR_{i,j}}
\quad\st\quad (\ref{nnoma1c}),
(\ref{nnoma1d}), (\ref{nnoma2b}), (\ref{nnoma2c}), (\ref{nnomatkf})-(\ref{nnomatkba}),
\end{eqnarray}
\else
\begin{eqnarray}\label{nnomatk2}
\ds\min_{\bw, \balpha, \clR}\
\frac{ \xi \pi(\bw)+P_c}{\sum_{(i,j)\in\clI\times\clJ}\ \clR_{i,j}} \quad\st \nonumber\\
(\ref{nnoma1c}), (\ref{nnoma1d}), (\ref{nnoma2b}), (\ref{nnoma2c})-(\ref{nnomatkh}), (\ref{nnomatkba}).
\end{eqnarray}
\fi
Algorithm \ref{alg4} outlines the steps to solve the PS-based algorithm for energy-efficiency (EE) maximization problem (\ref{nnoma2}). Like Algorithm \ref{alg3},  at least it converges to a locally optimal solution of \eqref{nnoma2}, which satisfies the KKT optimality condition.

Locating a feasible point $(\mathbf{w}^{(0)}, \clR^{(0)}, \balpha^{(0)})$ for the
nonconvex constraints (\ref{nnoma1b}), (\ref{R1}), (\ref{R2}), and
(\ref{R3}) to initialize Algorithm \ref{alg4} is resolved
via the iterations
\ifCLASSOPTIONpeerreview
\begin{eqnarray}\label{qos}
\ds\max_{\mathbf{w},\clR,\balpha} \min_{(i,j)\in\clI\times\clJ_c}\ \bigg\{\frac{\clR_{i,j}}{r_{i,j}}-1, \frac{\clR_{i,p(j)}}{r_{i,p(j)}}-1,
\ds\zeta_{i,j}\sum_{(s,\ell)\in\clS}\left(2\Re\{(\bw_{s,\ell}^{(\kappa)})^H\bh^H_{s,i,j}\bh_{s,i,j}\bw_{s,l}  \}
-|\bh_{s,i,j}\bw^{(\kappa)}_{s,l}|^2 \right)\nonumber\\
+\zeta_{i,j}\sigma^2-\frac{e^{\min}_{i,j}}{1-\alpha_{i,j}}  \bigg\}\quad
\st\quad \eqref{nnoma1c}, \eqref{nnoma1d},  \eqref{nnomatkb}-\eqref{nnomatkh}
\end{eqnarray}
\else
\begin{align}\label{qos}
\ds\max_{\mathbf{w},\clR,\balpha} \min_{(i,j)\in\clI\times\clJ_c}\ & \bigg\{\frac{\clR_{i,j}}{r_{i,j}}-1, \frac{\clR_{i,p(j)}}{r_{i,p(j)}}-1, \notag \\
& \hspace{-2.2cm} \ds\zeta_{i,j}\sum_{(s,\ell)\in\clS}\left(2\Re\{(\bw_{s,\ell}^{(\kappa)})^H\bh^H_{s,i,j}\bh_{s,i,j}\bw_{s,l}  \}
\right.\nonumber\\
& \hspace{-0.5cm} \left.-|\bh_{s,i,j}\bw^{(\kappa)}_{s,l}|^2 \right) +\zeta_{i,j}\sigma^2-\frac{e^{\min}_{i,j}}{1-\alpha_{i,j}}  \bigg\} \notag \\
\st &\quad \eqref{nnoma1c}, \eqref{nnoma1d},  \eqref{nnomatkb}-\eqref{nnomatkh}
\end{align}
\fi
till reaching a value more than or equal to $0$.

\begin{algorithm}[t]
\begin{algorithmic}[1]
\protect\caption{PS-based algorithm for energy-efficiency (EE) maximization problem (\ref{nnoma2})}
\label{alg4}
\global\long\def\algorithmicrequire{\textbf{Initialization:}}
\REQUIRE  Set $\kappa:=0$ and initialize a feasible point $(\bw^{(0)}, \clR^{(0)}, \balpha^{(0)})$ for (\ref{nnoma2}).
\STATE Until convergence of the objective in (\ref{nnoma2}) repeat: Solve the convex  optimization problem (\ref{nnomatk2}) to obtain the optimal solution $\mathbf{w}^{(\kappa+1)},\balpha^{(\kappa+1)},\clR^{(\kappa+1)}$.
Reset $\kappa:=\kappa+1.$
\end{algorithmic} \end{algorithm}
\section{Transmit TS-based NOMA}
In a time-switching (TS) based system, a fraction of time $0< \rho < 1$ is used for power transfer while the remaining fraction of time $1-\rho$ for information transfer, where $\rho$ is termed as TS ratio. For power transfer, we have to design the energy beamforming vectors $\mathbf{w}^E_{i,j}$, $\forall$, $i \in \mathcal{I}$, $j \in \mathcal{J}_c$.
For $p_{i,j}(\mathbf{w}^E)\triangleq \sum_{(s,\ell)\in {\cal S}_c}|\mathbf{h}_{s,i,j}\mathbf{w}^E_{s,\ell}|^2$
and $\zeta_{i,j} \in (0,1)$ as the energy conversion efficiency for the EH receiver,
the harvested energy by the cell-center user UE$(i,j)$ is expressed as
\begin{align}
   E_{i,j}(\mathbf{w}^E,\rho) &\triangleq \rho \zeta_{i,j}
   ( p_{i,j}(\mathbf{w}^E) + \sigma^2).
\end{align}
Here, we assume a common TS ratio $\rho$ for all BSs, $i \in \mathcal{I}$, where near-by users harvest energy through wireless signals not only from the serving BSs but also from the neighboring BSs. Note that the harvested and stored energy $E_{i,j}$ may be used later for different power constrained operations at cell-center UEs $(i,j)$, e.g., assisting uplink data transmission to the BS or performing downlink information processing. Let us denote $\mathbf{w}^E \triangleq [\mathbf{w}^E_{i,j}]_{i\in{\cal I}, j\in {\cal J}_{c}}$.

The remaining time ($1-\rho$) will be used for information decoding by all users UE $(i,j)$, $i\in{\cal I}, j\in {\cal J}$. Let $\mathbf{w}^I \triangleq [\mathbf{w}^I_{i,j}]_{i\in{\cal I}, j\in {\cal J}}$ define the information beamforming vectors and suppose that $\clR_{i,p(j)}$ is the achievable rate by decoding
 $ s_{i,p(j)}$. Recalling the definitions (\ref{eqLMijb}), (\ref{eqLMipijb}), (\ref{lcij}) and (\ref{basef}), we have
\begin{eqnarray}
 \clR_{i,p(j)} \leq (1-\rho)\psi(\mathbf{h}_{i,i,j}\mathbf{w}^I_{i, p(j)},\clL_{i,p(j)}^c(\mathbf{w}^I),0),
            \label{R1T}
\end{eqnarray}
\begin{eqnarray}
\clR_{i,p(j)}\leq (1-\rho) \psi(\mathbf{h}_{i,i,p(j)}\mathbf{w}^I_{i,p(j)}, \clL_{i,p(j)}^e(\mathbf{w}^I), 0).\label{R2T}
 \end{eqnarray}
After decoding $s_{i,p(j)}$, UE $(i,j)$,  $j\in {\cal J}_c$ decodes $s_{i,j}$. The achievable rate $\clR_{i,j}$ by decoding $s_{i,j}$ is
 \begin{eqnarray}
\clR_{i,j}\leq (1-\rho)\psi(\mathbf{h}_{i,i,j}\mathbf{w}^I_{i,j},\clL_{i,j}^c(\mathbf{w}^I), 0). \label{R3T}
 \end{eqnarray}
Thus, the throughput maximin optimization problem for TS-based NOMA SWIPT system is given by
\ifCLASSOPTIONpeerreview
\begin{subequations}\label{nnoma1T}
\begin{eqnarray}
\ds\max_{\bw^I, \bw^E, \clR,\rho}\ \min_{(i,j)\in\clI\times\clJ}\ \clR_{i,j}\quad\st\quad (\ref{R1T})-(\ref{R3T}), \label{nnoma1aT}\\
\ds\zeta_{i,j} \left(p_{i,j}(\mathbf{w}^E) + \sigma^2\right)\geq \frac{e^{\min}_{i,j}}{\rho}, i\in\clI, j\in\clJ_c,\label{nnoma1bT}\\
0<\rho <1, \label{nnoma1cT}\\
\frac{1}{1-\rho}\sum_{j\in {\clJ}_c} \|\mathbf{w}^E_{i,j}\|^2+  \sum_{j\in \clJ} \|\mathbf{w}^I_{i,j}\|^2 \le
\ds\frac{P^{\max}_{i}}{1-\rho}+\sum_{j\in {\clJ}_c} \|\mathbf{w}^E_{i,j}\|^2 ,\ i \in\mathcal{I}\label{nnoma1dT}\\
\|\mathbf{w}^E_{i,j}\|^2    \le P^{\max}_{i},    i \in\mathcal{I}, j \in \mathcal{J}_c \label{nnoma1eT}\\
\|\mathbf{w}^I_{i,j}\|^2    \le P^{\max}_{i},    i \in\mathcal{I}, j \in \mathcal{J}. \label{nnoma1fT}
\end{eqnarray}
\end{subequations}
\else
\begin{subequations}\label{nnoma1T}
\begin{eqnarray}
\ds\max_{\bw^I, \bw^E, \clR,\rho}\ \min_{(i,j)\in\clI\times\clJ}\ \clR_{i,j}\quad\st\quad (\ref{R1T}), (\ref{R2T}), (\ref{R3T}), \label{nnoma1aT}\\
\ds\zeta_{i,j} \left(p_{i,j}(\mathbf{w}^E) + \sigma^2\right)\geq \frac{e^{\min}_{i,j}}{\rho}, i\in\clI, j\in\clJ_c,\label{nnoma1bT}\\
0<\rho <1, \label{nnoma1cT}\\
\frac{1}{1-\rho}\sum_{j\in {\clJ}_c} \|\mathbf{w}^E_{i,j}\|^2+  \sum_{j\in \clJ} \|\mathbf{w}^I_{i,j}\|^2 \le
 \notag \\ \ds\frac{P^{\max}_{i}}{1-\rho}+\sum_{j\in {\clJ}_c} \|\mathbf{w}^E_{i,j}\|^2 ,\  i \in\mathcal{I}\label{nnoma1dT}\\
\|\mathbf{w}^E_{i,j}\|^2    \le P^{\max}_{i}   \  i \in\mathcal{I}, j \in \mathcal{J}_c \label{nnoma1eT}\\
\|\mathbf{w}^I_{i,j}\|^2    \le P^{\max}_{i}   \  i \in\mathcal{I}, j \in \mathcal{J}. \label{nnoma1fT}
\end{eqnarray}
\end{subequations}
\fi
Note that the power constraint (\ref{nnoma1dT}) is the equivalent reexpression of the sum power constraint
\[
\rho \sum_{j\in {\clJ}_c} \|\mathbf{w}^E_{i,j}\|^2 +  (1-\rho) \sum_{j\in \clJ} \|\mathbf{w}^I_{i,j}\|^2 \le P^{\max}_{i},\ \forall i \in\mathcal{I}.
\]
{\color{black}Let $(\bw^{I, (\kappa)}, \bw^{E, (\kappa)}, \clR^{(\kappa)},\rho^{(\kappa)})$ be a feasible point
for (\ref{nnoma1T}) that is found from the $(\kappa-1)$th iteration. By using the inequality
$x\leq 0.5(x^2/\bar{x}+\bar{x})$ $\forall x), \bar{x}>0$ \cite[(78)]{Sheetal18} and (\ref{es1}), we can innerly approximate  (\ref{R1T}), (\ref{R2T}) and (\ref{R3T}) by
\ifCLASSOPTIONpeerreview
\begin{eqnarray}\label{R1Tk}
 \ds 0.5\frac{(\clR_{i,p(j)})^2/\clR^{(\kappa)}_{i,p(j)}+\clR^{(\kappa)}_{i,p(j)}}{1-\rho} &\leq&
 \Lambda_0(\mathbf{h}_{i,i,j}\mathbf{w}^I_{i, p(j)},\clL_{i,p(j)}^c(\mathbf{w}^I)),
\end{eqnarray}
\begin{eqnarray}\label{R12Tk}
 \ds 0.5\frac{(\clR_{i,p(j)})^2/\clR^{(\kappa)}_{i,p(j)}+\clR^{(\kappa)}_{i,p(j)}}{1-\rho} &\leq&
 \Lambda_0(\mathbf{h}_{i,i,p(j)}\mathbf{w}^I_{i, p(j)},\clL_{i,p(j)}^e(\mathbf{w}^I))
\end{eqnarray}
and
\begin{eqnarray}\label{R3Tk}
 \ds 0.5\frac{(\clR_{i,j})^2/\clR^{(\kappa)}_{i,j}+\clR^{(\kappa)}_{i,j}}{1-\rho} &\leq&
 \Lambda_0(\mathbf{h}_{i,i,j}\mathbf{w}^I_{i, j},\clL_{i,j}^c(\mathbf{w}^I))
\end{eqnarray}
\else
\begin{align}\label{R1Tk}
 \ds 0.5\frac{  \frac{(\clR_{i,p(j)})^2}{\clR^{(\kappa)}_{i,p(j)}}+\clR^{(\kappa)}_{i,p(j)}}{1-\rho} \leq
 \Lambda_0(\mathbf{h}_{i,i,j}\mathbf{w}^I_{i, p(j)},\clL_{i,p(j)}^c(\mathbf{w}^I)),
\end{align}
\begin{align}\label{R12Tk}
 \ds 0.5\frac{ \frac{(\clR_{i,p(j)})^2}{\clR^{(\kappa)}_{i,p(j)}}+\clR^{(\kappa)}_{i,p(j)}}{1-\rho} \leq
 \Lambda_0(\mathbf{h}_{i,i,p(j)}\mathbf{w}^I_{i, p(j)},\clL_{i,p(j)}^e(\mathbf{w}^I))
\end{align}
and
\begin{align}\label{R3Tk}
 \ds 0.5\frac{\frac{(\clR_{i,j})^2}{\clR^{(\kappa)}_{i,j}}+\clR^{(\kappa)}_{i,j}}{1-\rho} \leq
 \Lambda_0(\mathbf{h}_{i,i,j}\mathbf{w}^I_{i, j},\clL_{i,j}^c(\mathbf{w}^I))
\end{align}
\fi
over the trust regions
\ifCLASSOPTIONpeerreview
\begin{subequations}\label{trustts1}
\begin{eqnarray}
2\Re\{(\bwIk_{i,p(j)})^H\mathbf{h}^H_{i,i,p(j)} \mathbf{h}_{i,i,p(j)}\bw^I_{i,p(j)}\}- \|\mathbf{h}_{i,i,p(j)}\bwIk_{i,p(j)}\|^2
>0,\label{trustts1a}\\
2\Re\{(\bwIk_{i,p(j)})^H\mathbf{h}^H_{i,i,j}\mathbf{h}_{i,i,j}\bw^I_{i,p(j)}\}-\|\mathbf{h}_{i,i,j}\bwIk_{i,p(j)} \|^2
>0,\label{trustts1b}\\
2\Re\{(\bwIk_{i,j})^H\mathbf{h}^H_{i,i,j}\mathbf{h}_{i,i,j}\bw^I_{i,j}\}- \|\mathbf{h}_{i,i,j}\bwIk_{i,j}\|^2>0,\label{trustts1c}
\end{eqnarray}
\end{subequations}
\else
\begin{subequations}\label{trustts1}
\begin{eqnarray}
2\Re\{(\bwIk_{i,p(j)})^H\mathbf{h}^H_{i,i,p(j)} \mathbf{h}_{i,i,p(j)}\bw^I_{i,p(j)}\}&& \notag \\ - \|\mathbf{h}_{i,i,p(j)}\bwIk_{i,p(j)}\|^2
&>&0,\label{trustts1a}\\
2\Re\{(\bwIk_{i,p(j)})^H\mathbf{h}^H_{i,i,j}\mathbf{h}_{i,i,j}\bw^I_{i,p(j)}\} && \notag \\ -\|\mathbf{h}_{i,i,j}\bwIk_{i,p(j)} \|^2
&>&0,\label{trustts1b}\\
2\Re\{(\bwIk_{i,j})^H\mathbf{h}^H_{i,i,j}\mathbf{h}_{i,i,j}\bw^I_{i,j}\} &&\notag \\ - \|\mathbf{h}_{i,i,j}\bwIk_{i,j}\|^2&>&0,\label{trustts1c}
\end{eqnarray}
\end{subequations}
\fi
with  the concave functions $\Lambda_0(\mathbf{h}_{i,i,j}\mathbf{w}^I_{i, p(j)},\clL_{i,p(j)}^c(\mathbf{w}^I))$,
$\Lambda_0(\mathbf{h}_{i,i,p(j)}\mathbf{w}^I_{i, p(j)},\clL_{i,p(j)}^e(\mathbf{w}^I))$, and
$ \Lambda_0(\mathbf{h}_{i,i,j}\mathbf{w}^I_{i, j},\clL_{i,j}^c(\mathbf{w}^I)) $ defined from
(\ref{es1a}), (\ref{es2})-(\ref{es4}) for $(\bx^{(\kappa)},\bfy^{(\kappa)})=
(\mathbf{h}_{i,i,j}\mathbf{w}^{I, (\kappa)}_{i, p(j)},\clL_{i,p(j)}^c(\mathbf{w}^{I, (\kappa)}))$,
$(\bx^{(\kappa)},\bfy^{(\kappa)})=(\mathbf{h}_{i,i,p(j)}\mathbf{w}^{I, (\kappa)}_{i, p(j)},\clL_{i,p(j)}^e(\mathbf{w}^{I, (\kappa)}))$,
and  $(\bx^{(\kappa)},\bfy^{(\kappa)})=(\mathbf{h}_{i,i,j}\mathbf{w}^{I, (\kappa)}_{i, j},\clL_{i,j}^c(\mathbf{w}^{I, (\kappa)}))$, respectively.

The EH constraint (\ref{nnoma1bT}) is innerly approximated by \cite{Nasir-17-Nov-A}
\ifCLASSOPTIONpeerreview
\begin{eqnarray}
\ds  \sum_{(s,\ell)\in\mathcal{S}_c}  \left[2\Re\left\{ (\mathbf{w}^{E,(\kappa)}_{s,\ell} )^H \mathbf{h}_{s,i,j}^H
\mathbf{h}_{s,i,j} \mathbf{w}^E_{s,\ell}\right\} - \left|\mathbf{h}_{i,i,j} \mathbf{w}^{E,(\kappa)}_{s,\ell}\right|^2\right] + \sigma^2
\geq \ds\frac{e^{\min}_{i,j}}{\zeta_{i,j}\rho},    i \in \mathcal{I}, j \in \mathcal{J}_c, \label{c8b}
\end{eqnarray}
\else
\begin{align}
&\hspace{-0.3cm} \ds  \sum_{(s,\ell)\in\mathcal{S}_c}  \left[2\Re\left\{ (\mathbf{w}^{E,(\kappa)}_{s,\ell} )^H \mathbf{h}_{s,i,j}^H
\mathbf{h}_{s,i,j} \mathbf{w}^E_{s,\ell}\right\} \right. \notag \\ &
-\left. \left|\mathbf{h}_{i,i,j} \mathbf{w}^{E,(\kappa)}_{s,\ell}\right|^2\right] + \sigma^2
\geq \ds\frac{e^{\min}_{i,j}}{\zeta_{i,j}\rho},  i \in \mathcal{I}, j \in \mathcal{J}_c, \label{c8b}
\end{align}
\fi
while the power constraint (\ref{nnoma1dT}) is innerly approximated by \cite{Nasir-17-Nov-A}
\ifCLASSOPTIONpeerreview
\begin{eqnarray}
\ds \frac{1}{1-\rho}\sum_{j\in\mathcal{J}_c} \| \mathbf{w}^E_{i,j} \|^2
+\sum_{j\in\mathcal{J}} \| \mathbf{w}^I_{i,j} \|^2&\le&\ds
\left(\frac{2}{1-\rho^{(\kappa)}}-\frac{1-\rho}{(1-\rho^{(\kappa)})^2}\right)P_i^{\max}  \nonumber\\
&&\ds +\sum_{j\in\mathcal{J}}\left( 2\Re\left\{(\mathbf{w}^{E,(\kappa)}_{i,j})^H \mathbf{w}^E_{i,j}\right\}
-\|\mathbf{w}^{E,(\kappa)}_{i,j}\|^2\right), \ i \in \mathcal{I}. \label{c2am}
\end{eqnarray}
\else
\begin{align}
 \ds  \frac{1}{1-\rho}\sum_{j\in\mathcal{J}_c} \| \mathbf{w}^E_{i,j} \|^2
+\sum_{j\in\mathcal{J}} \| \mathbf{w}^I_{i,j} \|^2 &\le \notag \\  \ds
\left(\frac{2}{1-\rho^{(\kappa)}}-\frac{1-\rho}{(1-\rho^{(\kappa)})^2}\right)P_i^{\max} & \nonumber\\
\ds +\sum_{j\in\mathcal{J}}\left( 2\Re\left\{(\mathbf{w}^{E,(\kappa)}_{i,j})^H \mathbf{w}^E_{i,j}\right\}
-\|\mathbf{w}^{E,(\kappa)}_{i,j}\|^2\right).& \label{c2am}
\end{align}
\fi
In summary,  at the $\kappa$th iteration we solve the following convex optimization problem of computational complexity
$\mathcal{O} \left(  (2 K N (N_t + N_t/2 + 1) + 1)^3  \left( 9 KN +  N + 1\right) \right)$
 to generate the feasible point $(\bw^{I, (\kappa+1)}, \bw^{E, (\kappa+1)}, \clR^{(\kappa)},\rho^{(\kappa+1)})$
for (\ref{nnoma1T}):
\ifCLASSOPTIONpeerreview
\begin{equation}\label{eq:P2}
 \ds\max_{\mathbf{w}^E,\mathbf{w}^I,\clR, \rho} \min_{(i,j)\in\clI\times\clJ} \clR_{i,j} \quad \mbox{s.t} \quad
 (\ref{nnoma1cT}), (\ref{nnoma1eT}), (\ref{nnoma1fT}),
 (\ref{R1Tk})-(\ref{c2am}).
\end{equation}
\else
\begin{eqnarray}\label{eq:P2}
 \ds\max_{\mathbf{w}^E,\mathbf{w}^I,\clR, \rho} \min_{(i,j)\in\clI\times\clJ}\ \clR_{i,j}
 \quad\st \nonumber\\
 (\ref{nnoma1cT}), (\ref{nnoma1eT}), (\ref{nnoma1fT}),
 (\ref{R1Tk})-(\ref{c2am}).
\end{eqnarray}
\fi
Algorithm~\ref{alg5} outlines the steps to solve the ``transmit TS-based" algorithm for throughput max-min optimization problem \eqref{nnoma1T}. Like Algorithm \ref{alg3}, it converges at least to a local optimal solution satisfying the KKT condition. To find the initial feasible point $(\mathbf{w}^{E,(0)},\mathbf{w}^{I,(0)},\rho^{(0)},\clR^{(0)})$ of \eqref{nnoma1T}, we first fix $\rho^{(0)}$ and find $(\mathbf{w}^{E,(0)},\mathbf{w}^{I,(0)})$ by randomly generating $N_t \times 1$ complex vectors followed by their normalization to satisfy \eqref{nnoma1dT}, \eqref{nnoma1eT}, and \eqref{nnoma1fT}. We then find $\left(\mathbf{w}^{E,(0)},\mathbf{w}^{I,(0)},\clR^{(0)}\right)$ via iterations
\ifCLASSOPTIONpeerreview
\begin{subequations}\label{qos6}
\begin{eqnarray}
\ds\max_{\mathbf{w}^E,\mathbf{w}^I,\clR} \min_{(i,j)\in\clI\times\clJ_c}\ \bigg\{
\ds\zeta_{i,j}\sum_{(s,\ell)\in\clS_c}\left(2\Re\{(\bw_{s,\ell}^{E,(\kappa)})^H\bh^H_{s,i,j}\bh_{s,i,j}\bw^E_{s,l} \}\right.
\nonumber\\
\ds\left.-|\bh_{s,i,j}\bw^{E,(\kappa)}_{s,l}|^2 \right)+\zeta_{i,j}\sigma^2-\frac{e^{\min}_{i,j}}{\rho^{(0)}}   \bigg\}\quad
\st\quad (\ref{nnoma1eT}), (\ref{nnoma1fT}),\label{qos6a}\\
\rho^{(0)} \sum_{j\in {\clJ}_c} \|\mathbf{w}^E_{i,j}\|^2 +  (1-\rho^{(0)}) \sum_{j\in \clJ} \|\mathbf{w}^I_{i,j}\|^2 \le P^{\max}_{i},\ \forall i \in\mathcal{I}, \label{qos6b}\\
 \ds\frac{\clR_{i,p(j)}}{1-\rho^{(0)}} \leq
 \Lambda_0(\mathbf{h}_{i,i,j}\mathbf{w}^I_{i, p(j)},\clL_{i,p(j)}^c(\mathbf{w}^I)),\label{qos6c}\\
 \ds \frac{\clR_{i,p(j)}}{1-\rho^{(0)}} \leq
 \Lambda_0(\mathbf{h}_{i,i,p(j)}\mathbf{w}^I_{i, p(j)},\clL_{i,p(j)}^e(\mathbf{w}^I)), \label{qos6d}\\
 \ds \frac{\clR_{i,j}}{1-\rho^{(0)}} \leq
 \Lambda_0(\mathbf{h}_{i,i,j}\mathbf{w}^I_{i, j},\clL_{i,j}^c(\mathbf{w}^I)),\label{qos6e}
\end{eqnarray}
\end{subequations}
\else
\begin{subequations}\label{qos6}
\begin{align}
 \ds \max_{\mathbf{w}^E,\mathbf{w}^I,\clR} \min_{(i,j)\in\clI\times\clJ_c}\ & \notag \\ & \hspace{-2cm} \bigg\{
\ds\zeta_{i,j}\sum_{(s,\ell)\in\clS_c}\left(2\Re\{(\bw_{s,\ell}^{E,(\kappa)})^H\bh^H_{s,i,j}\bh_{s,i,j}\bw^E_{s,l}  \}\right.  \nonumber\\
& \hspace{-1.2cm}  \ds\left.-|\bh_{s,i,j}\bw^{E,(\kappa)}_{s,l}|^2 \right) +\zeta_{i,j}\sigma^2-\frac{e^{\min}_{i,j}}{\rho^{(0)}}  \bigg\}\notag \\
& \hspace{-2.5cm}\st\quad (\ref{nnoma1eT}), (\ref{nnoma1fT}),\label{qos6a}  \\
& \hspace{-2.2cm}  \rho^{(0)} \sum_{j\in {\clJ}_c} \|\mathbf{w}^E_{i,j}\|^2 +  (1-\rho^{(0)}) \sum_{j\in \clJ} \|\mathbf{w}^I_{i,j}\|^2 \le P^{\max}_{i}, \label{qos6b}   \displaybreak \\
&\hspace{-2cm}   \ds\frac{\clR_{i,p(j)}}{1-\rho^{(0)}} \leq
 \Lambda_0(\mathbf{h}_{i,i,j}\mathbf{w}^I_{i, p(j)},\clL_{i,p(j)}^c(\mathbf{w}^I)),\label{qos6c}\\
 &\hspace{-2cm}   \ds \frac{\clR_{i,p(j)}}{1-\rho^{(0)}} \leq
 \Lambda_0(\mathbf{h}_{i,i,p(j)}\mathbf{w}^I_{i, p(j)},\clL_{i,p(j)}^e(\mathbf{w}^I)), \label{qos6d}\\
& \hspace{-2cm}   \ds \frac{\clR_{i,j}}{1-\rho^{(0)}} \leq
 \Lambda_0(\mathbf{h}_{i,i,j}\mathbf{w}^I_{i, j},\clL_{i,j}^c(\mathbf{w}^I)),\label{qos6e}
\end{align}
\end{subequations}
\fi
till reaching a value more than or equal to $0$, for a fixed $\rho^{(0)}$.
If problem \eqref{qos6} is infeasible with $\rho^{(0)}$ or solving \eqref{qos6} fails to give a positive optimal value, we repeat the above process for a different value of $\rho^{(0)}$ in order to find a feasible point $\left(\mathbf{w}^{E,(0)}, \mathbf{w}^{I,(0)},\clR^{(0)}, \rho^{(0)}\right)$ for \eqref{nnoma1T}.\footnote{Simulation results in Sec.~\ref{sec:sim} show that in almost all of the scenarios considered, problem \eqref{qos6} is feasible and a positive optimal value of \eqref{qos6} is obtained in a just two iterations for the first tried value $\rho^{(0)} = 0.2$.}

\begin{algorithm}[t]
\begin{algorithmic}[1]
\protect\caption{Transmit TS-based algorithm for throughput max-min optimization problem \eqref{nnoma1T}}
\label{alg5}
\global\long\def\algorithmicrequire{\textbf{Initialization:}}
\REQUIRE  Set $\kappa:=0$ and initialize a feasible point $\left(\bw^{E,(0)},\bw^{I, (0)},\clR^{(0)},\rho^{(0)} \right)$ for \eqref{nnoma1T}.
\STATE Until convergence of the objective in \eqref{nnoma1T} repeat: Solve the convex  optimization problem (\ref{eq:P2}) to obtain the optimal solution $\left(\mathbf{w}^{E,(\kappa+1)},\mathbf{w}^{I,(\kappa+1)}, \clR^{(\kappa + 1)}, \rho^{(\kappa+1)} \right)$.
Reset $\kappa:=\kappa+1$.
\end{algorithmic} \end{algorithm}
Next, we address the following energy-efficiency maximization problem
\ifCLASSOPTIONpeerreview
\begin{eqnarray}\label{ee1}
\ds\max_{\mathbf{w}^E,\mathbf{w}^I,\clR, \rho}\
\frac{\sum_{(i,j)\in\clI\times\clJ}\ \clR_{i,j}}{\xi \left[\rho\pi_E(\bw^E)+(1-\rho)\pi_I(\bw^I) \right]+P_c}\nonumber\\
\st\quad (\ref{nnoma2b}), (\ref{nnoma2c}),
 (\ref{R1T})-(\ref{R3T}), (\ref{nnoma1bT})-(\ref{nnoma1fT})
\end{eqnarray}
\else
\begin{eqnarray}\label{ee1}
\ds\max_{\mathbf{w}^E,\mathbf{w}^I,\clR, \rho} \
\frac{\sum_{(i,j)\in\clI\times\clJ}\ \clR_{i,j}}{\xi \left[\rho\pi_E(\bw^E)+(1-\rho)\pi_I(\bw^I) \right]+P_c}\
\st\nonumber\\ (\ref{nnoma2b}), (\ref{nnoma2c}),
 (\ref{R1T})-(\ref{R3T}), (\ref{nnoma1bT})-(\ref{nnoma1fT})
\end{eqnarray}
\fi
where $\pi_E(\bw^E)=\sum_{i\in\clI}\sum_{j\in {\clJ}_c} \|\mathbf{w}^E_{i,j}\|^2$
and $\pi_I(\bw^I)=\sum_{i\in\clI}\sum_{j\in \clJ} \|\mathbf{w}^I_{i,j}\|^2$.
 Note that we define $\rho\pi_E(\bw^E)+(1-\rho)\pi_I(\bw^I)$ to differentiate from $\pi(\bw)$ in
(\ref{nnoma2e}). For this network, it is more appropriate
 to use $(1-\rho)\pi_I(\bw^I)$ for information delivery.

In contrast to the objective function in (\ref{nnoma2}), the power consumption function in the denominator of the
objective function in (\ref{ee1}) is no longer convex, making equivalently transformed problem like (\ref{nnoma2})
no longer useful. Nevertheless, we now also develop another iterative procedure, where (\ref{ee1}) is seen no
more computationally difficult than the throughput optimization problem  (\ref{nnoma1T}).

Let  $(\bw^{E, (\kappa)}, \bw^{I, (\kappa)},\clR^{(\kappa)}, \rho^{(\kappa)})$ be a feasible point for (\ref{ee1}) that is found from the $(\kappa-1)$th iteration and
\ifCLASSOPTIONpeerreview
\[
\begin{array}{lll}
t^{(\kappa)}&\triangleq&\ds\frac{\sum_{(i,j)\in\clI\times\clJ}\ \clR^{(\kappa)}_{i,j}}{\xi \left[ \rho^{(\kappa)}\pi_E(\bw^{E, (\kappa)})+(1-\rho^{(\kappa)})\pi_I(\bw^{I, (\kappa)})\right] +P_c}\nonumber\\
&=&\ds\frac{\sum_{(i,j)\in\clI\times\clJ}\ \clR^{(\kappa)}_{i,j}/(1-\rho^{(\kappa)})}{ \xi
(1/(1-\rho^{(\kappa)})-1)\pi_E(\bw^{E, (\kappa)})+ \xi \pi_I(\bw^{I, (\kappa)})+P_c/(1-\rho^{(\kappa)})}
\end{array}
\]
\else
\[
\begin{array}{rl}
t^{(\kappa)}\triangleq&\ds \frac{\sum_{(i,j)\in\clI\times\clJ}\ \clR^{(\kappa)}_{i,j}}{\xi \left[ \rho^{(\kappa)}\pi_E(\bw^{E, (\kappa)})+(1-\rho^{(\kappa)})\pi_I(\bw^{I, (\kappa)})\right] +P_c}\nonumber\\
=&\ds\frac{\sum_{(i,j)\in\clI\times\clJ}\ \clR^{(\kappa)}_{i,j}/(1-\rho^{(\kappa)})}{ \xi
(\frac{\pi_E(\bw^{E, (\kappa)})}{(1-\rho^{(\kappa)})-1} + \xi \pi_I(\bw^{I, (\kappa)})+\frac{P_c}{1-\rho^{(\kappa)}}}
\end{array}
\]
\fi
At the $\kappa$th iteration we address the problem
\ifCLASSOPTIONpeerreview
\begin{eqnarray}\label{ee3}
\ds\max_{\mathbf{w}^E,\mathbf{w}^I,\clR, \rho}\
\frac{\sum_{(i,j)\in\clI\times\clJ}\ \clR_{i,j}}{1-\rho}-t^{(\kappa)}\left[
\xi \left(\frac{1}{1-\rho}-1 \right)\pi_E(\bw^E)+\xi \pi_I(\bw^I)+\frac{P_c}{1-\rho}\right]\nonumber\\
\st\quad
(\ref{nnoma2b}), (\ref{nnoma2c}),
(\ref{R1T})-(\ref{R3T}), (\ref{nnoma1bT})-(\ref{nnoma1fT}).
\end{eqnarray}
\else
\begin{eqnarray}\label{ee3}
\max_{\mathbf{w}^E,\mathbf{w}^I,\clR, \rho}\
\frac{\sum_{(i,j)\in\clI\times\clJ}\ \clR_{i,j}}{1-\rho}- t^{(\kappa)}\left[
\xi \left(\frac{1}{1-\rho}-1 \right)\right.  \notag \\ \left. \times \pi_E(\bw^E)+\xi \pi_I(\bw^I)+\frac{P_c}{1-\rho}\right]\quad\st\nonumber\\
(\ref{nnoma2b}), (\ref{nnoma2c}),
(\ref{R1T})-(\ref{R3T}), (\ref{nnoma1bT})-(\ref{nnoma1fT}).
\end{eqnarray}
\fi
Substituting $x=\sum_{(i,j)\in\clI\times\clJ} \clR_{i,j}$, $\bar{x}=\sum_{(i,j)\in\clI\times\clJ} \clR^{(\kappa)}_{i,j}$,
and $t=1-\rho$, $\bar{t}=1-\rho^{(\kappa)}$ into the following inequality
\begin{equation}\label{newin}
\frac{x}{t}\geq 2\frac{\sqrt{\bar{x}}}{\bar{t}}\sqrt{x}-\frac{\bar{x}}{\bar{t}^2}t\ \quad\forall\ x>0, \bar{x}>0, t>0, \bar{t}>0
\end{equation}
whose proof is given by Appendix B,
the first term in the objective of (\ref{ee3}) is lower bounded  by the concave function
\ifCLASSOPTIONpeerreview
\[
f^{(\kappa)}(\clR,\rho)\triangleq
2\frac{\sqrt{\sum_{(i,j)\in\clI\times\clJ}\ \clR^{(\kappa)}_{i,j}}}{1-\rho^{(\kappa)}}
\sqrt{\sum_{(i,j)\in\clI\times\clJ}\ \clR_{i,j}}
-\frac{\sum_{(i,j)\in\clI\times\clJ}\ \clR^{(\kappa)}_{i,j}}{(1-\rho^{(\kappa)})^2}(1-\rho),
\]
\else
\[
\begin{array}{ll}
f^{(\kappa)}(\clR,\rho) \triangleq&  \ds
2\frac{\sqrt{\sum_{(i,j)\in\clI\times\clJ}\ \clR^{(\kappa)}_{i,j}}}{1-\rho^{(\kappa)}}
\sqrt{\sum_{(i,j)\in\clI\times\clJ}\ \clR_{i,j}}\\
&-\ds\frac{\sum_{(i,j)\in\clI\times\clJ}\ \clR^{(\kappa)}_{i,j}}{(1-\rho^{(\kappa)})^2}(1-\rho),
\end{array}
\]
\fi
{\color{black}The second term  $g(\bw,\rho) \triangleq
\xi \left(\frac{1}{1-\rho}-1 \right)\pi_E(\bw^E)+\xi \pi_I(\bw^I)+\frac{P_c}{1-\rho}$
in the objective of (\ref{ee3}) is upper bounded} by the convex function
\ifCLASSOPTIONpeerreview
\[
g^{(\kappa)}(\bw,\rho)\triangleq \frac{1}{(1-\rho)} \xi\pi_E(\bw^E)
+\xi\pi_I(\bw^I)+\frac{P_c}{1-\rho}- \xi
\sum_{i\in\clI}\sum_{j\in {\clJ}_c}\left(2\Re\{(\mathbf{w}^{E,(\kappa)}_{i,j})^H \mathbf{w}^E_{i,j} \} -\|\mathbf{w}^{E,(\kappa)}_{i,j}\|^2 \right),
\]
\else
\[
\begin{array}{r}
g^{(\kappa)}(\bw,\rho)\triangleq \frac{1}{(1-\rho)} \xi\pi_E(\bw^E)
 +\xi\pi_I(\bw^I)+\frac{P_c}{1-\rho} \\ - \xi
\sum_{i\in\clI}\sum_{j\in {\clJ}_c}\left(2\Re\{(\mathbf{w}^{E,(\kappa)}_{i,j})^H \mathbf{w}^E_{i,j} \} -\|\mathbf{w}^{E,(\kappa)}_{i,j}\|^2 \right),
\end{array}
\]
\fi
because $\|\bx\|^2 \ge 2 \Re\{(\bxk)^H\bx\}-\|\bxk\|^2$.

Thus, we solve the following convex optimization problem of computational complexity $\mathcal{O} \left(   (2 K N (N_t + N_t/2 + 1) + 1)^3  \left( 11 KN +  N + 1\right) \right)$
to generate $(\bw^{E,(\kappa+1)},\bw^{I,(\kappa+1)},\clR^{(\kappa+1)}, \rho^{(\kappa+1)})$
\ifCLASSOPTIONpeerreview
\begin{eqnarray}\label{ee4}
\ds\max_{\mathbf{w}^E,\mathbf{w}^I,\clR, \rho}\ [ f^{(\kappa)}(\clR,\rho)-
t^{(\kappa)}g^{(\kappa)}(\bw,\rho)]\nonumber\\
 \mbox{s.t} \quad
(\ref{nnoma2b}), (\ref{nnoma2c}),
 (\ref{R1Tk})-(\ref{c2am}).
\end{eqnarray}
\else
\begin{eqnarray}\label{ee4}
\ds\max_{\mathbf{w}^E,\mathbf{w}^I,\clR, \rho}\ [ f^{(\kappa)}(\clR,\rho)-
t^{(\kappa)}g^{(\kappa)}(\bw,\rho)]\quad\st\nonumber\\
(\ref{nnoma2b}), (\ref{nnoma2c}),
 (\ref{R1Tk})-(\ref{c2am}).
\end{eqnarray}
\fi
Algorithm \ref{alg6} outlines the steps to solve the ``transmit-TS based" algorithm for the energy-efficiency (EE) maximization problem (\ref{ee1}).

To initialize Algorithm \ref{alg6}, locating a feasible point $\left(\bw^{E,(0)},\bw^{I,(0)},\rho^{(0)},\clR^{(0)} \right)$ for (\ref{ee1}) is resolved via iterations
\ifCLASSOPTIONpeerreview
\begin{eqnarray}\label{qos6A4}
\ds\max_{\mathbf{w}^E,\mathbf{w}^I,\clR,\rho} \min_{(i,j)\in\clI\times\clJ_c}\ \bigg\{\frac{\clR_{i,j}}{r_{i,j}}-1, \frac{\clR_{i,p(j)}}{r_{i,p(j)}}-1 \bigg\} \nonumber\\
 \mbox{s.t} \quad
(\ref{nnoma2b}), (\ref{nnoma2c}),
 (\ref{R1Tk})- (\ref{R3Tk}), (\ref{trustts1}), (\ref{c8b}), (\ref{c2am}).
\end{eqnarray}
\else
\begin{eqnarray}\label{qos6A4}
\ds\max_{\mathbf{w}^E,\mathbf{w}^I,\clR,\rho} \min_{(i,j)\in\clI\times\clJ_c}\ \bigg\{\frac{\clR_{i,j}}{r_{i,j}}-1, \frac{\clR_{i,p(j)}}{r_{i,p(j)}}-1 \bigg\}\quad \mbox{s.t}  \nonumber\\
(\ref{nnoma2b}), (\ref{nnoma2c}),
 (\ref{R1Tk})- (\ref{R3Tk}),  (\ref{trustts1}), (\ref{c8b}), (\ref{c2am}).
\end{eqnarray}
\fi
till reaching a value greater than or equal to $0$. Its feasible point $\left(\bw^{E,(0)},\bw^{I,(0)},\rho^{(0)},\clR^{(0)} \right)$ for initialization is resolved via iterations \eqref{qos6}.}
\begin{algorithm}[t]
\begin{algorithmic}[1]
\protect\caption{Transmit TS-based algorithm for energy-efficiency (EE) maximization problem \eqref{ee1}}
\label{alg6}
\global\long\def\algorithmicrequire{\textbf{Initialization:}}
\REQUIRE  Set $\kappa:=0$ and initialize a feasible point $ \left(\bw^{E,(0)},\bw^{I,(0)}, \rho^{(0)}, \clR^{(0)} \right)$ for \eqref{ee1}.
\STATE Until convergence of the objective in \eqref{ee1} repeat: Solve the convex  optimization problem (\ref{ee4}) to obtain the optimal solution $ \left(\mathbf{w}^{E,(\kappa+1)},\mathbf{w}^{I,(\kappa+1)}, \clR^{(\kappa + 1)}, \rho^{(\kappa+1)} \right)$.
Reset $\kappa:=\kappa+1$.
\end{algorithmic} \end{algorithm}

\section{Simulation Results}\label{sec:sim}

To analyze the proposed algorithms through simulations,  a network topology as shown in Fig.~\ref{fig:nw_top} is set up.
There are $N = 3$ cells and $2K = 4$ UEs per cell with two
 placed close to the BS and the remaining two placed near cell-edges.
  {\color{black}These $12$ users are served over the same channel (same time and frequency) while we assume that other users will be allocated different frequency band or different time for communication.} The cell radius is set to be $100$ meters, where, near-by users are placed about the distance of $10$ meters from the serving BS while cell-edge users are placed about the distance of $80-90$ meters from the serving BS.

\ifCLASSOPTIONpeerreview
 \begin{figure}[t]
    \centering
    \includegraphics[width=0.65 \textwidth]{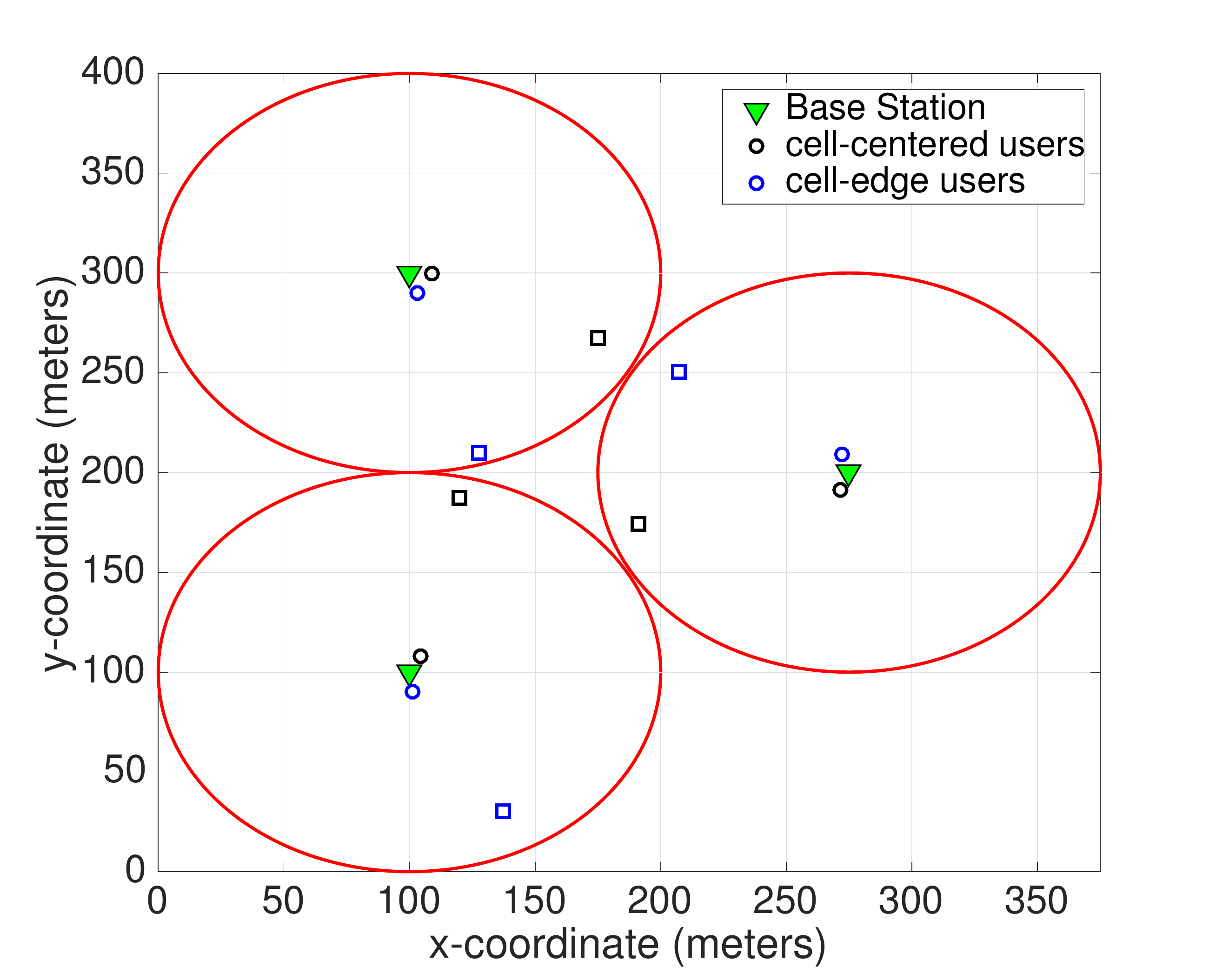}
  \caption{A multicell network setup used in our numerical examples.}
  \label{fig:nw_top}
\end{figure}
\else
 \begin{figure}[t]
    \centering
    \includegraphics[width=0.47 \textwidth]{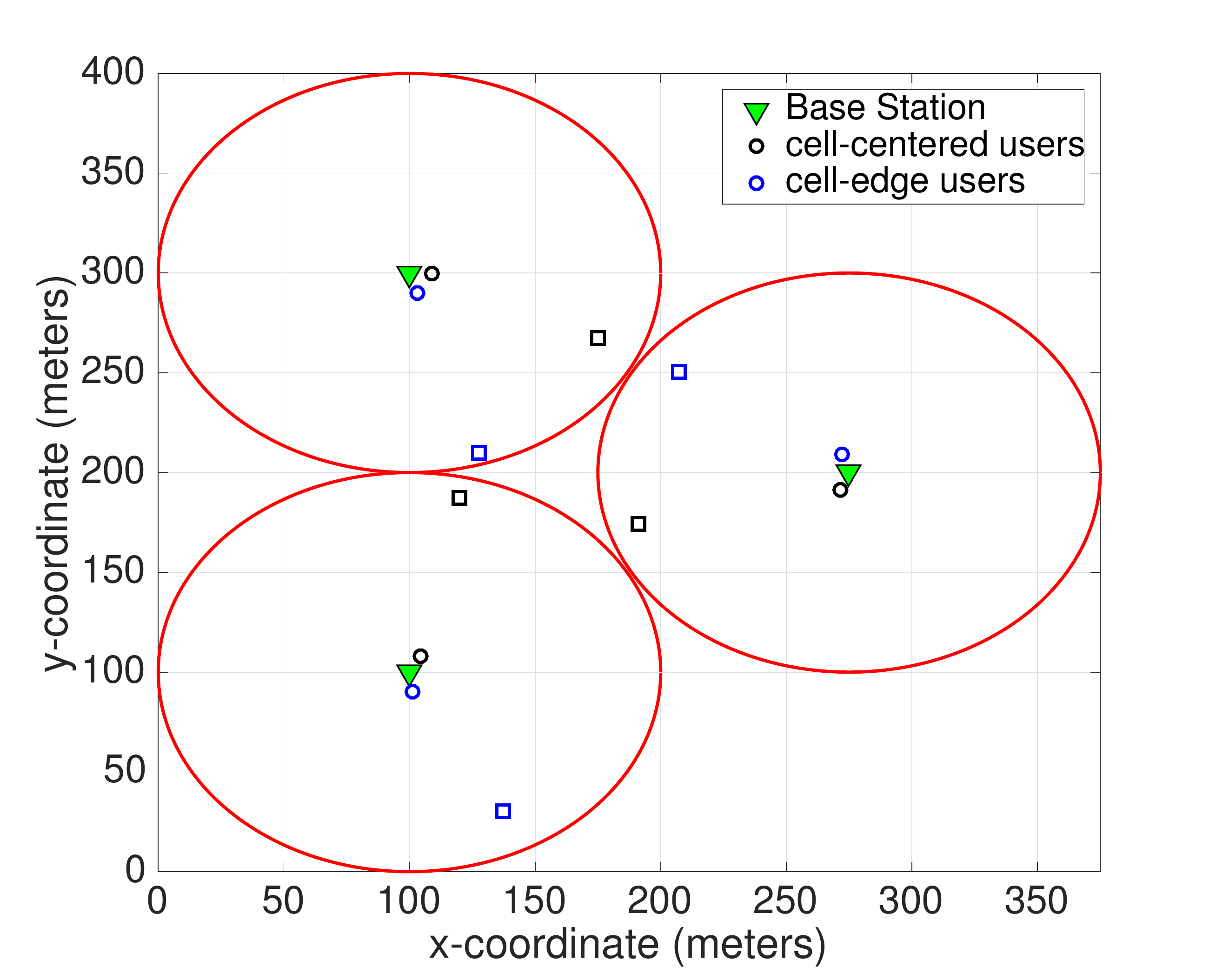}
  \caption{A multicell network setup used in our numerical examples.}
  \label{fig:nw_top}
\end{figure}
\fi

The channel $\mathbf{h}_{s,i,j}$  from BS $s \in \mathcal{I}$ to UE $(i,j)$ at a distance of $d$ meters is generated as $\mathbf{h}_{s,i,j} = \sqrt{10^{-\sigma_\text{PL}/10}} \tilde{\mathbf{h}}_{s,i,j}$, where $\sigma_\text{PL} = 30 + 10 \beta \log_{10} (d)$ is the path-loss in dB (the path-loss model is consistent with dense user deployment settings and wireless energy harvesting requirements \cite{Do-17-A-EA}), $\beta$ is the path-loss exponent, and $\tilde{\mathbf{h}}_{s,i,j}$ is the normalized Rayleigh fading channel gain (for $s = i$ while $j \in \mathcal{J}_e$ or $s \ne i$, i.e., channel between BS and its own cell-edge users or channel between BS and users in the neigboring cells) or $\tilde{\mathbf{h}}_{s,i,j}$ is the Rician fading channel gain with Rician factor of $10$ dB (for $s = i$ while $j \in \mathcal{J}_c$, i.e., channel between BS and its own cell-centered users). We set the path-loss exponents $\beta = 3$ for the former case and $\beta = 2$ for the later case. Different values for path-loss exponents have been proposed for different type of users in the literature too \cite{Liu-17-Dec-A,Ghazanfari-Apr-16-A}. For simplicity, \color{black}set $e_{i,j}^{\min} \equiv e^{\min}$ for the energy harvesting thresholds, $\zeta_{i,j} \equiv \zeta$, $\ \forall i,j$ for the energy harvesting conversion, $P_i^\text{max} \equiv P^\text{max}$, $\forall \ i$. Further, we set the energy harvesting threshold $e^{\min} = -20$ dBm and energy conversion efficiency $\zeta = 0.5$, $P^\text{max} = 35$ dBm (unless stated otherwise), noise variances $\sigma^2 = \sigma_c^2 =  -174$ dBm/Hz (unless stated otherwise), bandwidth = $20$ MHz, and carrier frequency = $2$ GHz. For energy efficiency maximization problems \eqref{nnoma2} and \eqref{ee1}, we set the threshold rate  $r_{i,j} = 0.5$ bits/sec/Hz $\forall \ i,j$ (unless stated otherwise), we choose the power amplifier efficiency $1/\xi = 0.2$, power dissipation at each transmit antenna $P_A = 0.6$W ($27.78$  dBm), and circuit power consumption $P_\text{cir} = 2.5$W ($33.97$ dBm)  \cite{Imran-11-EPD-A,Leng-15-ICNC-P}.

\subsection{Results for throughput max-min optimization problems \eqref{nnoma1}  and \eqref{nnoma1T}:}
On average (running $100$ simulations and averaging over random channel realizations), the PS-based Algorithm \ref{alg3} requires $16$ iterations, while TS-based Algorithm \ref{alg5} requires $19$ iterations before convergence.

\begin{figure*}[t]
    \centering
    \begin{minipage}[h]{0.47\textwidth}
    \centering
    \includegraphics[width=1.01 \textwidth]{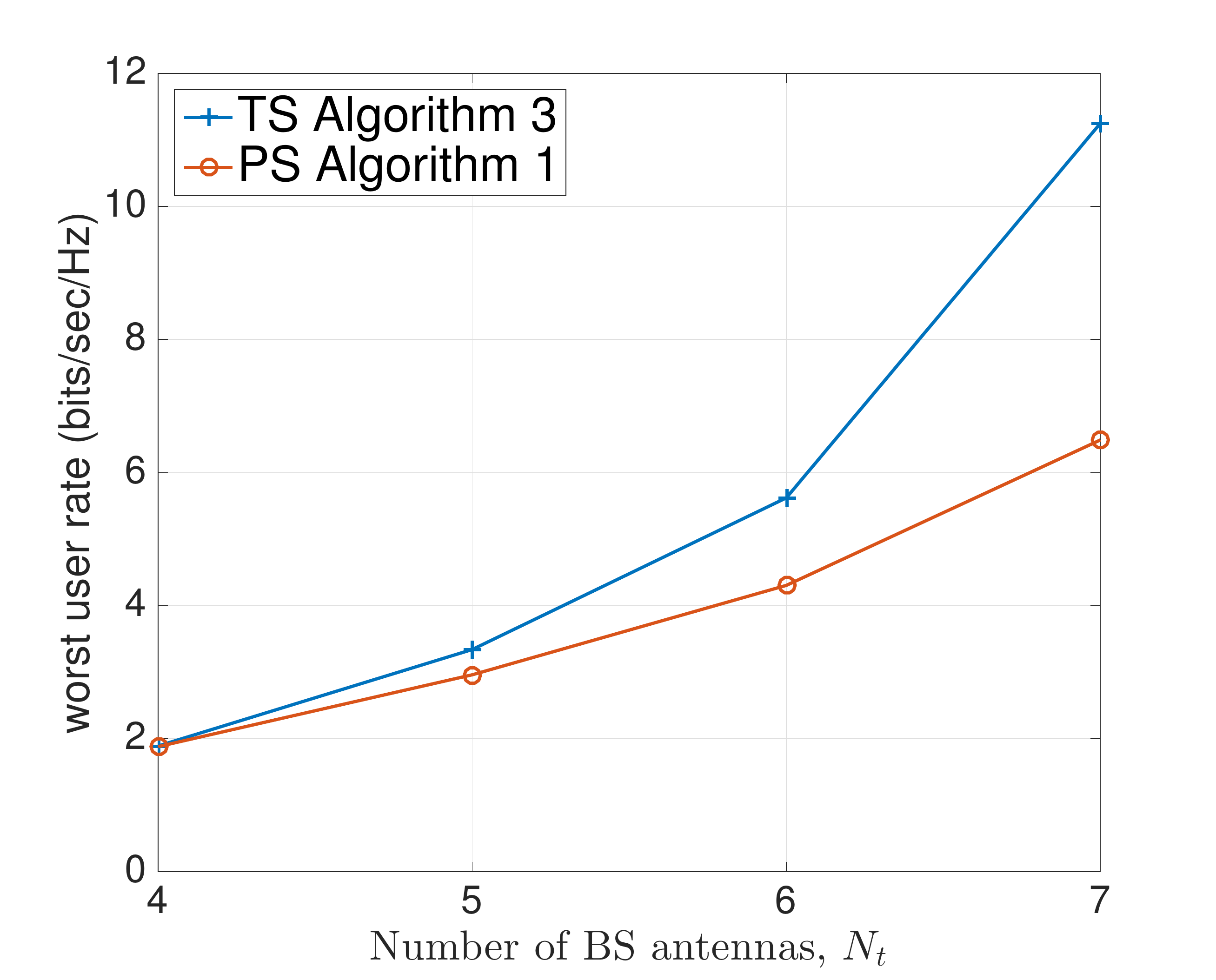}
  \caption{Optimized worst user rate for varying number of antennas and fixed BS power budget $P^\text{max} = 35$ dBm, while solving power splitting (PS)-based problem \eqref{nnoma1} and time switching (TS)-based problem \eqref{nnoma1T}.}
  \label{fig:mmR_M}
  \end{minipage}
    \hspace{0.3cm}
    \begin{minipage}[h]{0.47\textwidth}
    \centering
    \includegraphics[width=1.01 \textwidth]{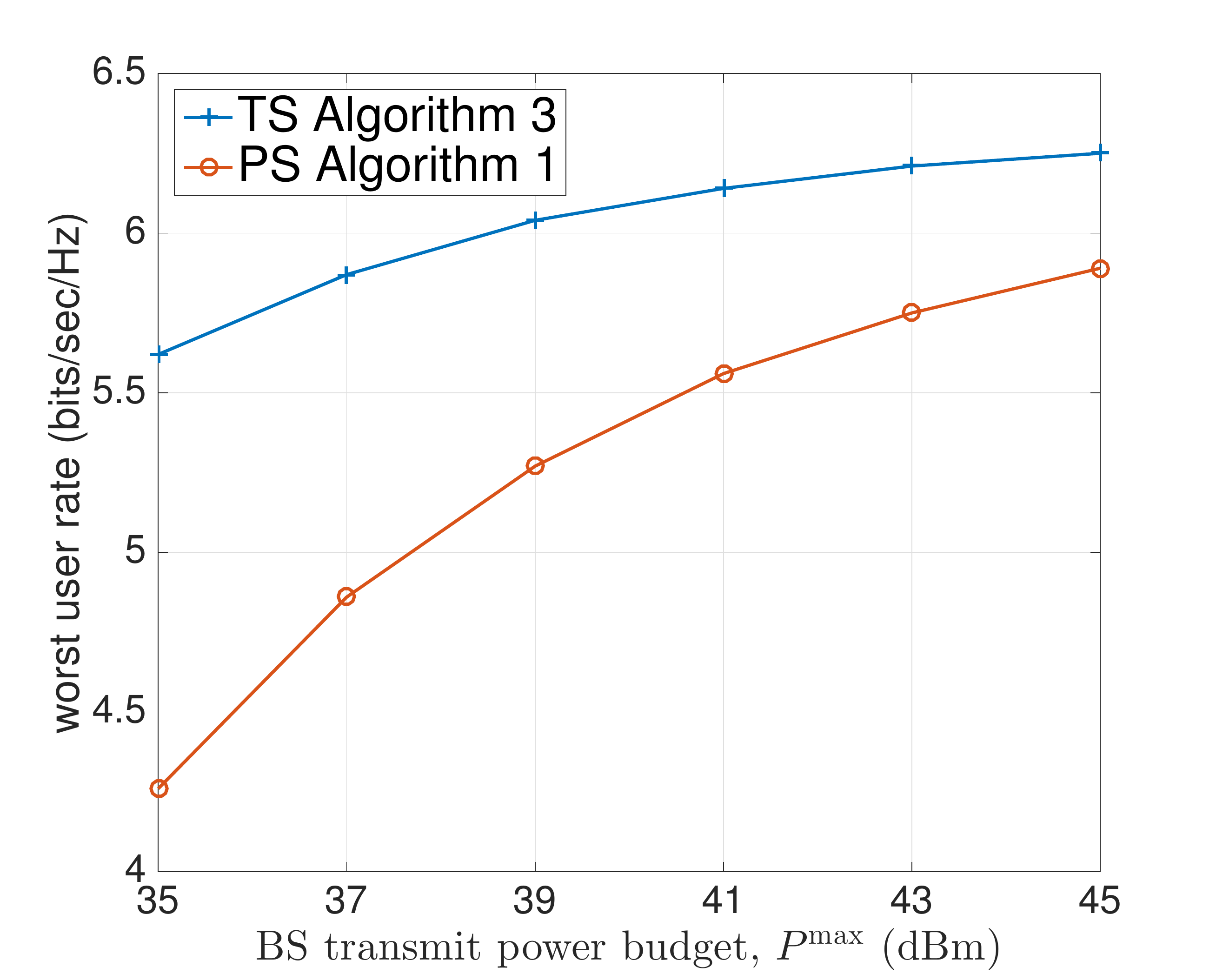}
  \caption{Optimized worst user rate for varying values of BS transmit power budget $P^\text{max}$ and fixed value of $N_t = 6$, while solving power splitting (PS)-based problem \eqref{nnoma1} and time switching (TS)-based problem \eqref{nnoma1T}.}
  \label{fig:mmR_P}
  \end{minipage}
\end{figure*}

\ifCLASSOPTIONpeerreview
\else
\begin{figure}[t]
    \centering
    \includegraphics[width=0.47 \textwidth]{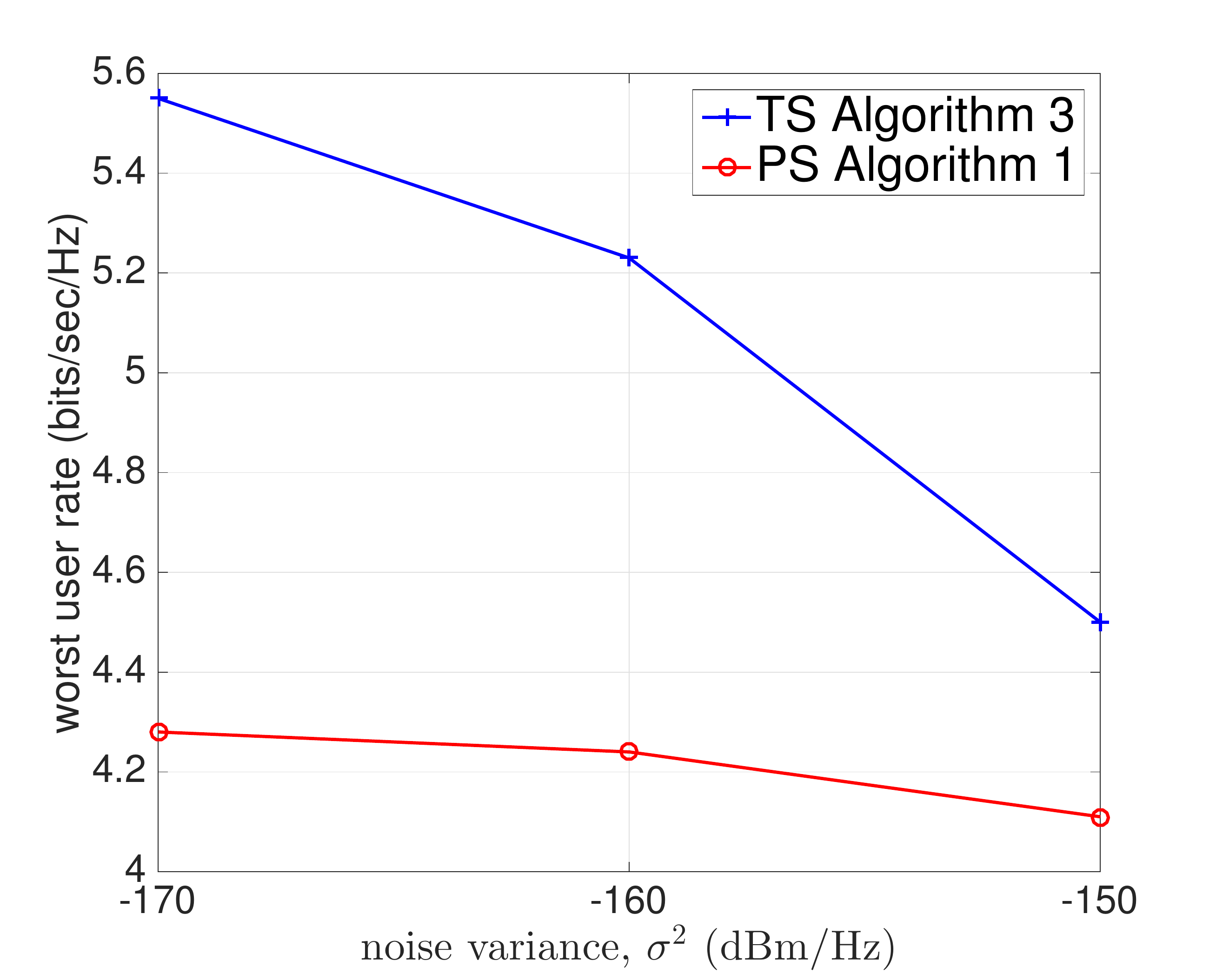}
  \caption{Optimized worst user rate for varying values of noise variances $\sigma^2$ and fixed value of $P^\text{max} = 35$ dBm and BS antennas  $N_t = 6$, while solving power splitting (PS)-based problem \eqref{nnoma1} and time switching (TS)-based problem \eqref{nnoma1T}.}
  \label{fig:mmR_N}
\end{figure}
\fi

Fig. \ref{fig:mmR_M} plots the optimized max-min rate for varying number of BS-antennas and fixed BS power budget $P^\text{max} = 35$ dBm, while solving the power splitting (PS)-based problem \eqref{nnoma1} and time switching (TS)-based problem \eqref{nnoma1T}. As expected, the rate increases by increasing the number of antennas at the BS. We can also observe from Fig. \ref{fig:mmR_M} that the TS-based Algorithm \ref{alg5} outperforms the PS-based Algorithm \ref{alg3} in terms of achievable rate and the corresponding performance gap increases if we increase the number of antennas mounted on the BS. {\color{black}This is because in the TS-based model, the presence of more antennas helps both information and energy beamforming vectors to scale their performance more progressively as compared to the PS-based model, where we only optimize information beamforming vectors.} Fig. \ref{fig:mmR_P} plots the optimized worst user rate for varying values of BS transmit power budget $P^\text{max}$ and fixed number of BS-antennas $N_t = 6$, while solving the same power splitting (PS)-based problem \eqref{nnoma1} and time switching (TS)-based problem \eqref{nnoma1T}. Fig. \ref{fig:mmR_P} shows that increasing the transmit power budget raises the level of achievable rate, however, the increase diminishes at higher values of transmit power budget, e.g., we can see marginal improvement in the rate when we increase the transmit power budget from $P^\text{max} = 43$ dBm to $45$ dBm. Similar to previous result in Fig. \ref{fig:mmR_M}, we observe from Fig. \ref{fig:mmR_P} that the TS-based Algorithm \ref{alg5} outperforms the PS-based Algorithm \ref{alg3} in terms of achievable rate.
\ifCLASSOPTIONpeerreview
\begin{figure}[t]
    \centering
    \includegraphics[width=0.65 \textwidth]{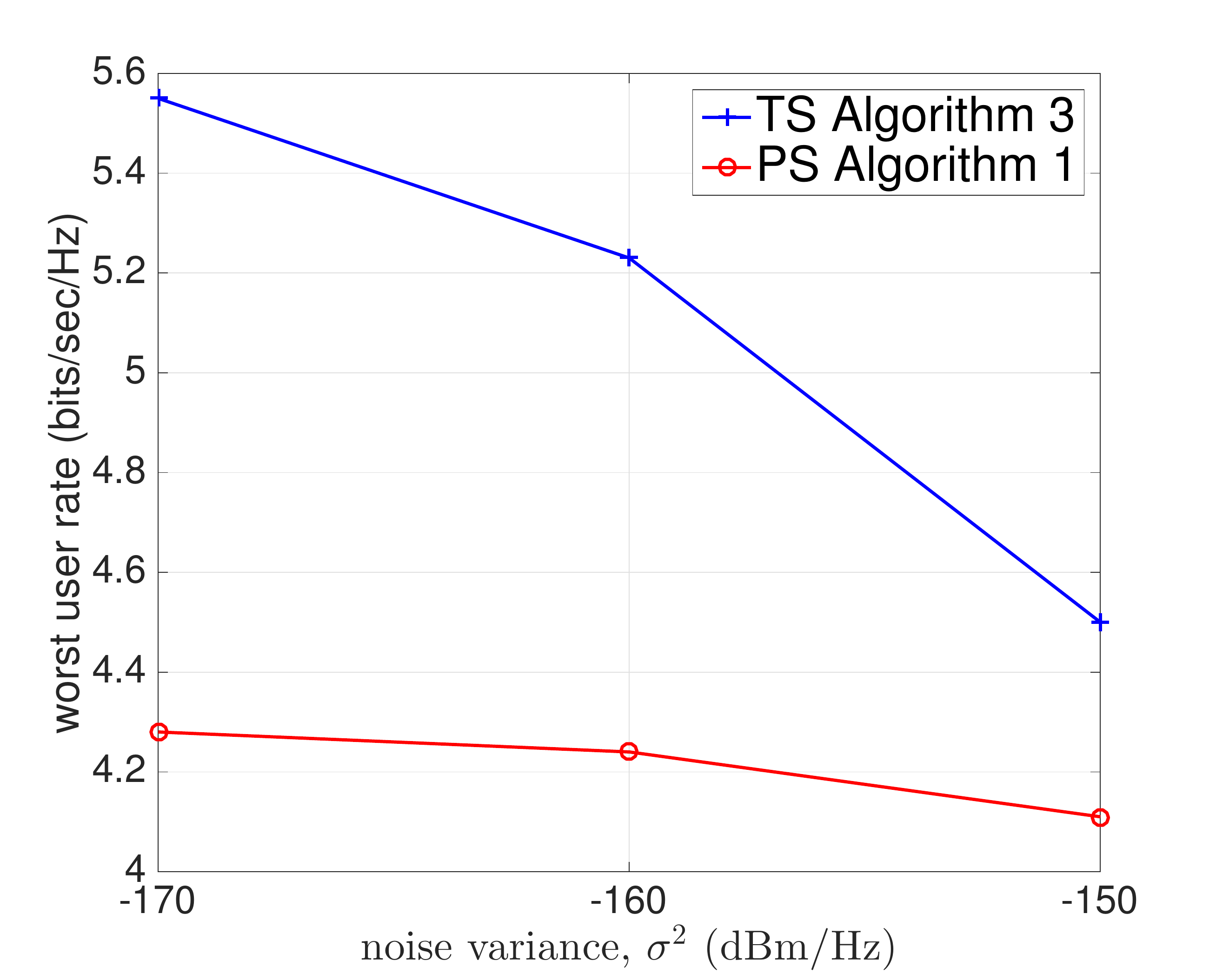} 
  \caption{Optimized worst user rate for varying values of noise variances $\sigma^2$ and fixed value of $P^\text{max} = 35$ dBm and BS antennas  $N_t = 6$, while solving the power splitting (PS)-based problem \eqref{nnoma1} and time switching (TS)-based problem \eqref{nnoma1T}.}
    \label{fig:mmR_N}
\end{figure}
\else
\fi
Fig. \ref{fig:mmR_N} plots the optimized worst user rate for varying values of noise variances $\sigma^2$ (in dBm/Hz) and fixed value of transmit power budget $P^\text{max} = 35$ dBm and BS antennas  $N_t = 6$, while solving the same power splitting (PS)-based problem \eqref{nnoma1} and time switching (TS)-based problem \eqref{nnoma1T}. Fig. \ref{fig:mmR_N} shows that, as expected, increasing the noise variance decreases the level of achievable rate. However, we observe a minor decrease in the achievable rate for PS-based receiver. Therefore, though the TS-based Algorithm \ref{alg5} outperforms the PS-based Algorithm \ref{alg3} in terms of achievable rate, but the corresponding performance gap decreases if we increase the noise variance.

{\color{black}
\begin{remark}
We observe from Figs. \ref{fig:mmR_M}-\ref{fig:mmR_N} that the TS-based Algorithm \ref{alg5} outperforms the PS-based Algorithm \ref{alg3} in terms of spectral efficiency. This is because for the TS-based Algorithm \ref{alg5}, we separately optimize the beamforming vectors for information transmission and energy harvesting, which results in better design compared to the case of PS-based Algorithm \ref{alg3}, where same beamforming vector is used for both information transmission and energy harvesting.
\end{remark}}
\subsection{Results for energy efficiency maximization problems \eqref{nnoma2} and \eqref{ee1}:}
On average, the PS-based Algorithm \ref{alg4} requires $13$ iterations, while the TS-based Algorithm \ref{alg6} requires $16$ iterations before convergence.

\begin{figure*}[t]
    \centering
    \begin{minipage}[h]{0.47\textwidth}
    \centering
    \includegraphics[width=1.01 \textwidth]{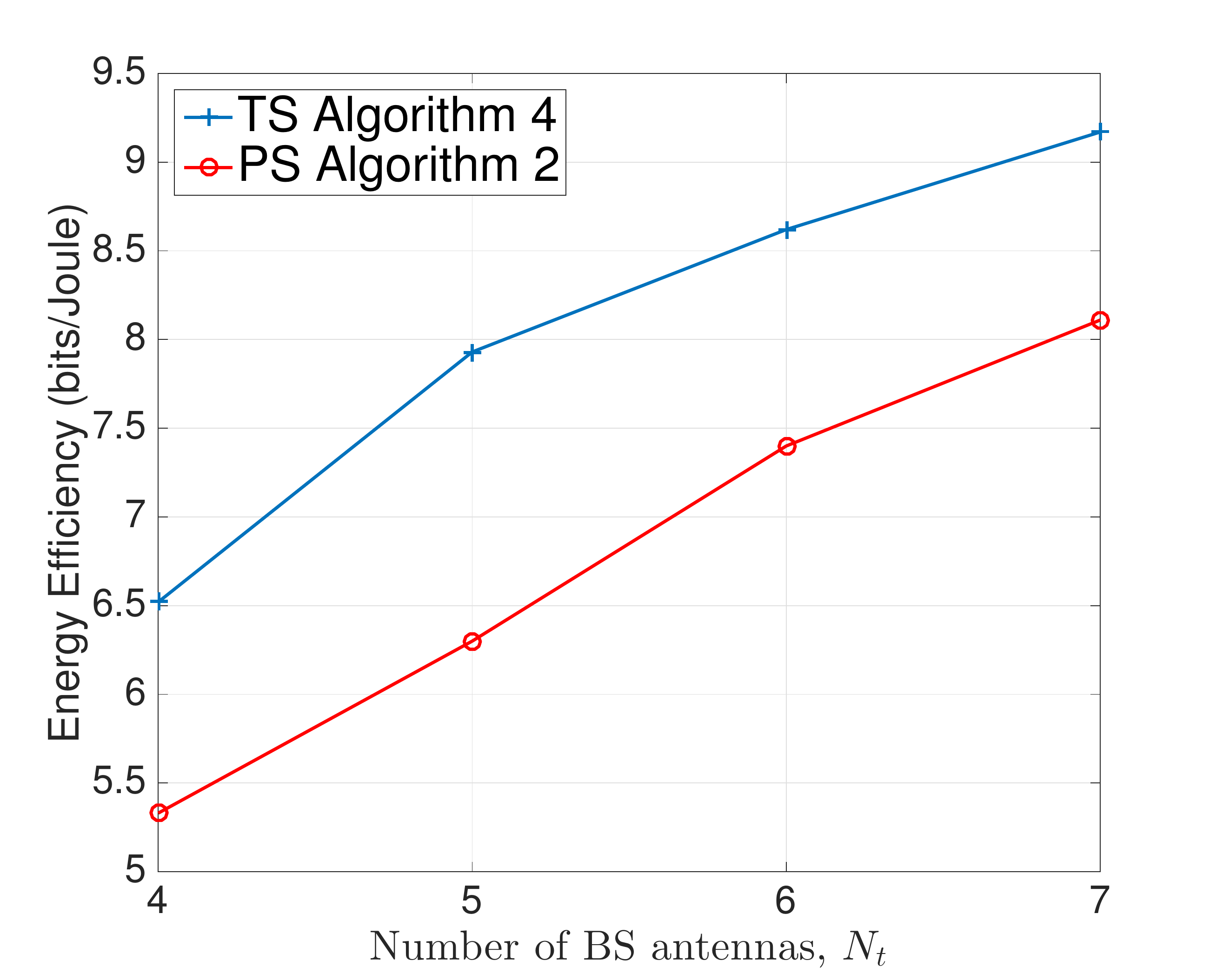}
  \caption{Optimized energy efficiency for varying number of antennas and fixed BS power budget $P^\text{max} = 35$ dBm, while solving power splitting (PS)-based problem \eqref{nnoma2} and time switching (TS)-based problem \eqref{ee1}.}
  \label{fig:EE_M}
  \end{minipage}
    \hspace{0.3cm}
    \begin{minipage}[h]{0.47\textwidth}
    \centering
    \includegraphics[width=1.01 \textwidth]{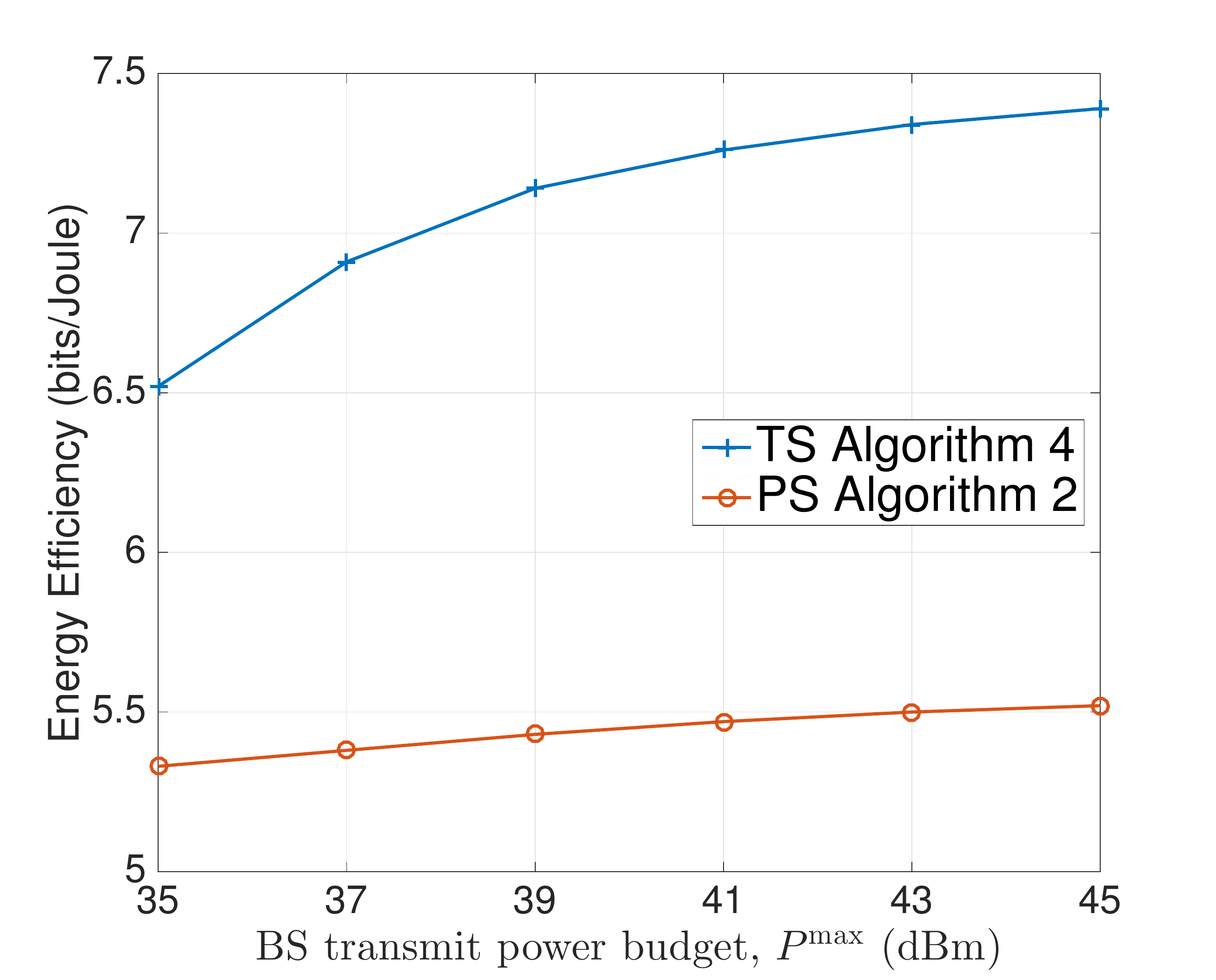}
  \caption{Optimized energy efficiency for varying values of noise variances $\sigma^2$ and fixed BS transmit power budget $P^\text{max} =  35$ and fixed value of $N_t = 4$, while solving power splitting (PS)-based problem \eqref{nnoma2} and time switching (TS)-based problem \eqref{ee1}}
  \label{fig:EE_P}
  \end{minipage}
\end{figure*}

\ifCLASSOPTIONpeerreview
\else
\begin{figure}[t]
    \centering
    \includegraphics[width=0.47 \textwidth]{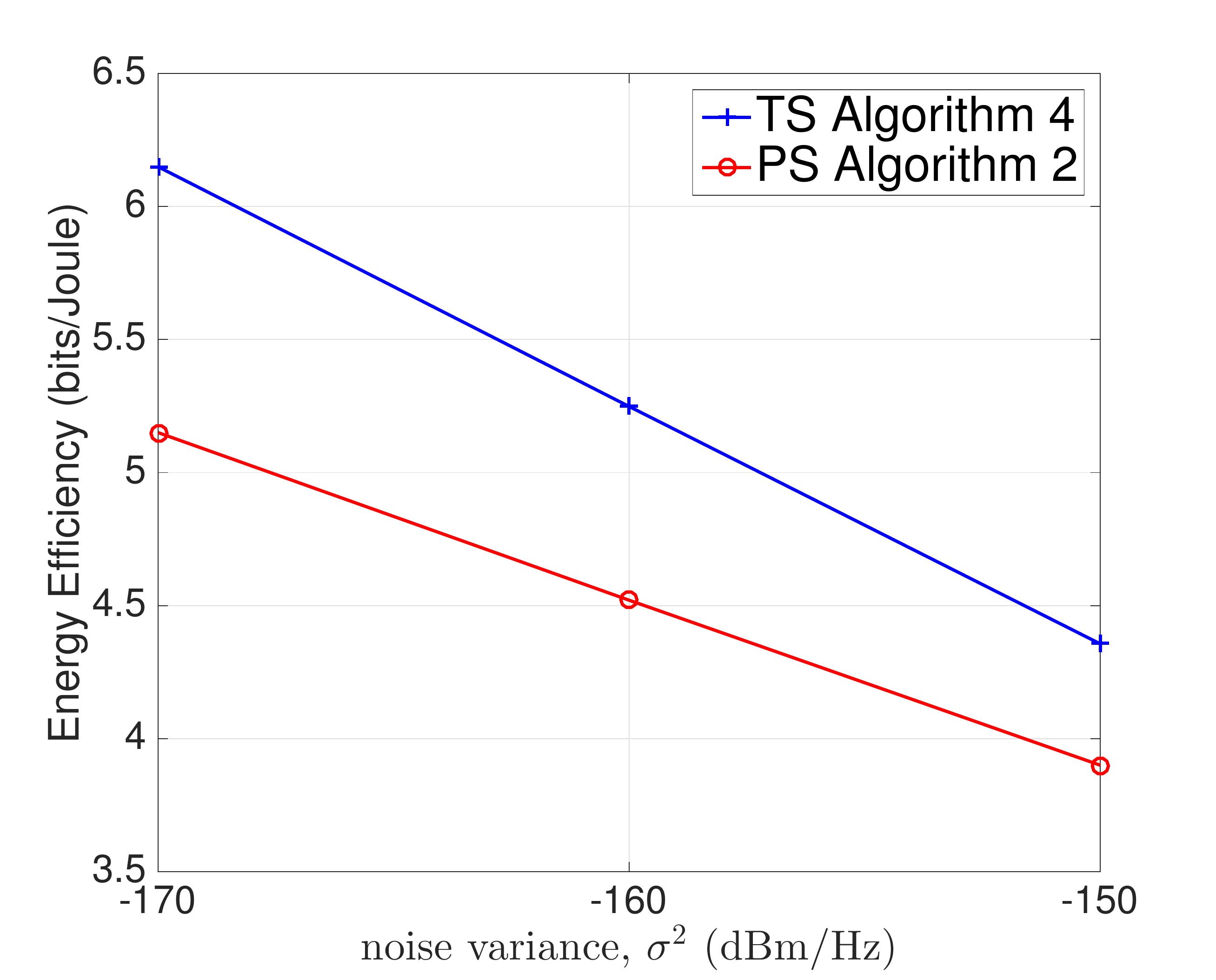} 
  \caption{Optimized energy efficiency for varying values of of noise variances $\sigma^2$ and fixed value of $P^\text{max} = 35$ dBm and BS antennas  $N_t = 4$, while solving power splitting (PS)-based problem \eqref{nnoma2} and time switching (TS)-based problem \eqref{ee1}}
    \label{fig:EE_N}
\end{figure}
\fi

Fig. \ref{fig:EE_M} plots the optimized energy efficiency (EE) for a varying number of BS-antennas and fixed BS power budget $P^\text{max} = 35$ dBm, while solving power splitting (PS)-based problem \eqref{nnoma2} and time switching (TS)-based problem \eqref{ee1}. As expected, the EE increases by increasing the number of antennas at the BS. We can also observe from Fig. \ref{fig:EE_M} that the TS-based Algorithm \ref{alg6} outperforms the PS-based Algorithm \ref{alg4} in terms of achievable EE and the performance gap is more than $1$ bit/Joule. Next, Fig. \ref{fig:EE_P} plots the energy efficiency for varying values of BS transmit power budget $P^\text{max}$ and fixed number of BS-antennas $N_t = 4$, while solving the same power splitting (PS)-based problem \eqref{nnoma2} and time switching (TS)-based problem \eqref{ee1}. Fig. \ref{fig:EE_P} shows that for TS-based implementation, increasing the transmit power budget raises the level of achievable EE, however, for PS-based implementation, there is almost no improvement in EE when the transmit power budget is increased from $P^\text{max} = 35$ dBm to $45$ dBm.

\ifCLASSOPTIONpeerreview
\begin{figure}[t]
    \centering
    \includegraphics[width=0.65 \textwidth]{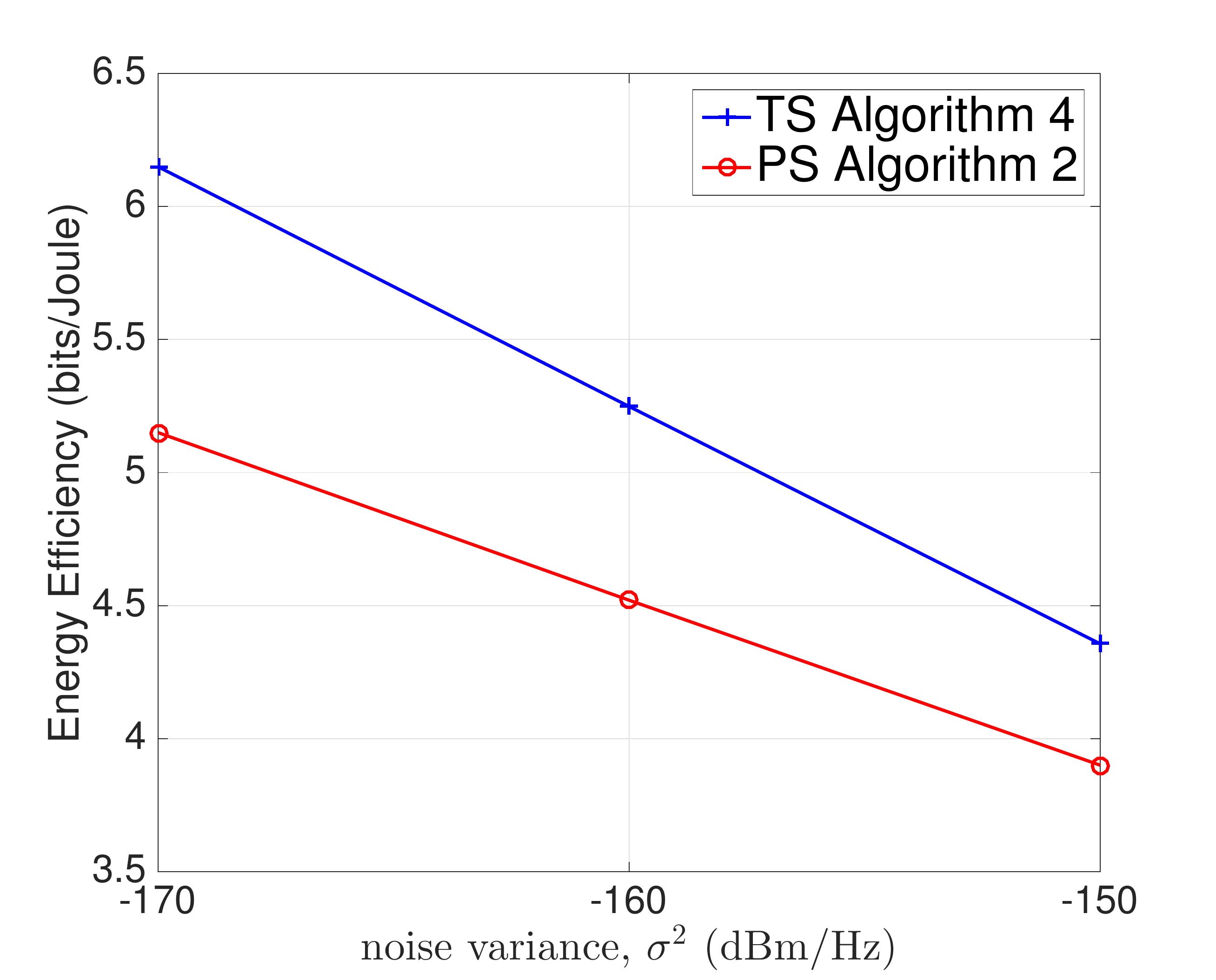} 
  \caption{Optimized energy efficiency for varying values of noise variances $\sigma^2$ and fixed value of $P^\text{max} = 35$ dBm and BS antennas  $N_t = 4$, while solving power splitting (PS)-based problem \eqref{nnoma2} and time switching (TS)-based problem \eqref{ee1}.}
    \label{fig:EE_N}
\end{figure}
\else
\fi

Fig. \ref{fig:EE_N} plots the optimized energy efficiency for different values of noise variances $\sigma^2$ (in dBm/Hz) and fixed value of transmit power budget $P^\text{max} = 35$ dBm and BS antennas  $N_t = 4$, while solving the power splitting (PS)-based problem \eqref{nnoma2} and time switching (TS)-based problem \eqref{ee1}. Fig. \ref{fig:EE_N} shows that, as expected, EE decreases by increasing the noise variance. In addition, the TS-based Algorithm \ref{alg6} outperforms the PS-based Algorithm \ref{alg4} in terms of EE. Nevertheless, this performance gap decreases with the increase of noise variance.

{\color{black}
\begin{remark}
We observe from Figs. \ref{fig:EE_M}-\ref{fig:EE_N} that the TS-based Algorithm \ref{alg6} outperforms the PS-based Algorithm \ref{alg4} in terms of energy efficiency. This is because for the TS-based Algorithm \ref{alg6}, we separately optimize the beamforming vectors for information transmission and energy harvesting, which results in better design compared to the case of PS-based Algorithm \ref{alg4}, where same beamforming vector is used for both information transmission and energy harvesting.
\end{remark}
}

\subsection{Comparison with orthogonal multiple access (OMA):}

\begin{figure*}[t]
    \centering
    \begin{minipage}[h]{0.47\textwidth}
    \centering
    \includegraphics[width=1.01 \textwidth]{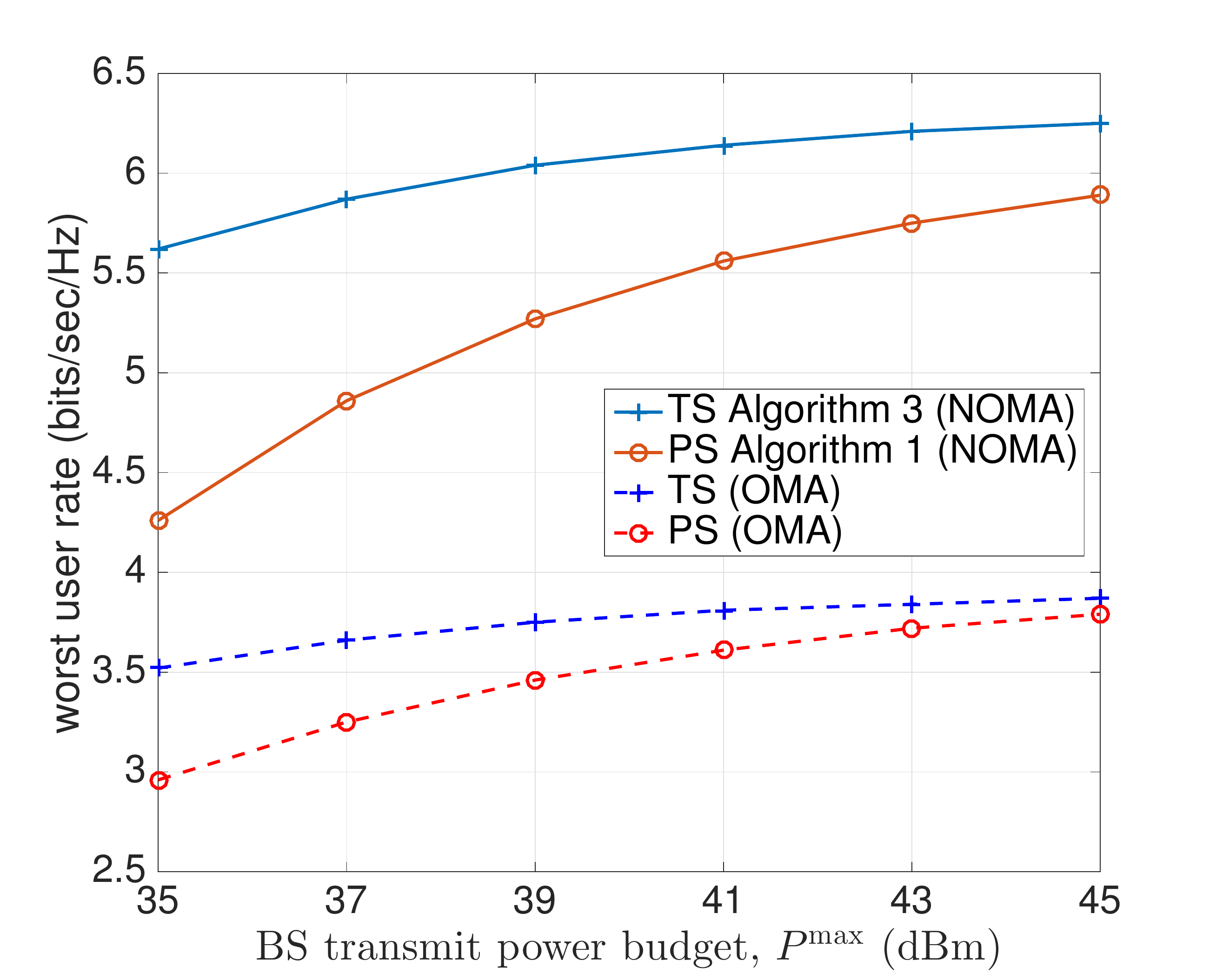}
  \caption{Optimized worst user rate for varying values of BS transmit power budget $P^\text{max}$ and fixed value of $N_t = 6$, while solving power splitting (PS)-based NOMA problem \eqref{nnoma1}, time switching (TS)-based NOMA problem \eqref{nnoma1T}, and their OMA counterparts.}
  \label{fig:mmR_P_OMA}
  \end{minipage}
    \hspace{0.3cm}
    \begin{minipage}[h]{0.47\textwidth}
    \centering
    \includegraphics[width=1.01 \textwidth]{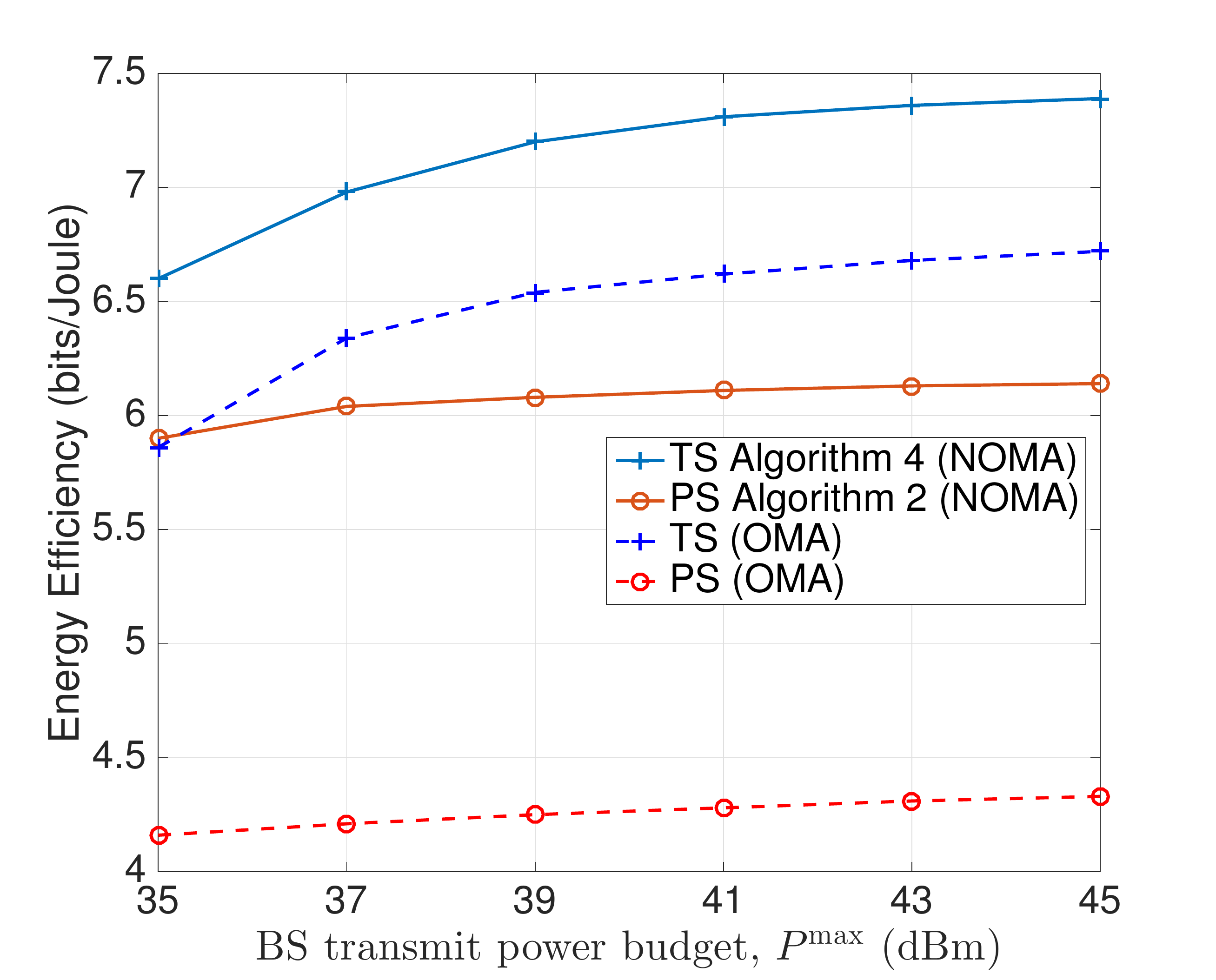}
  \caption{Optimized energy efficiency for varying values of BS transmit power budget $P^\text{max}$ and fixed value of $N_t = 4$ and threshold rate  $r_{i,j} = 0.1$ bits/sec/Hz (different from previous EE plots), while solving power splitting (PS)-based NOMA problem \eqref{nnoma2}, time switching (TS)-based NOMA problem \eqref{ee1}, and their OMA counterparts.}
  \label{fig:EE_P_OMA}
  \end{minipage}
\end{figure*}

Fig. \ref{fig:mmR_P_OMA} plots the optimized worst user rate for varying values of BS transmit power budget $P^\text{max}$ and fixed number of BS-antennas $N_t = 6$, while solving same power splitting (PS)-based NOMA problem \eqref{nnoma1}, time switching (TS)-based NOMA problem \eqref{nnoma1T}, and their orthogonal multiple access (OMA) counterparts. In parallel, Fig. \ref{fig:EE_P_OMA} plots the energy efficiency for varying values of BS transmit power budget $P^\text{max}$ and fixed number of BS-antennas $N_t = 4$ and threshold rate  $r_{i,j} = 0.1$ bits/sec/Hz (different from previous EE plots), while solving the power splitting (PS)-based NOMA problem \eqref{nnoma2}, time switching (TS)-based NOMA problem \eqref{ee1}, and their OMA counterparts. We have to choose a smaller threshold rate $r_{i,j} = 0.1$ bits/sec/Hz because implementation with OMA scheme fails to simultaneously satisfy both the higher threshold rate and the EH constraint \emph{for all simulations}. We can clearly observe from Figs. \ref{fig:mmR_P_OMA} and \ref{fig:EE_P_OMA} that NOMA implementation outperforms OMA implementation, in terms of both, throughput and energy efficiency, respectively.

\section{Conclusions}\label{sec:conclusion}

In this paper, we have considered energy harvesting based NOMA system, where transmit-TS approach is employed to realize wireless energy harvesting and information decoding at the nearly-located users. We have formulated two important problems of worst-user throughput maximization and energy efficiency maximization under power constraint and energy harvesting constraints at the nearly-located users. For these problems, the optimization objective and energy harvesting constraints are highly non-convex. To address this, we have developed efficient path-following algorithms to solve the two problems. We have also proposed algorithms for the case if conventional PS-based approach is used for energy harvesting. Our numerical results confirmed that the proposed transmit-TS approach clearly outperforms PS approach in terms of both, throughput and energy efficiency.


\section*{Appendix A: Proof for \eqref{es1}}
{\color{black}
Define the function $f(x,y) \triangleq \ln(x^{-1}+y^{-1})$ which is convex in $x>0$ and $y>0$ \cite{Taetal17}. Then
$f(x,y)\geq f(\xk,\yk)+\la \nabla f(\xk,\yk), (x,y)-(\xk,\yk)\ra$ for all $x>0$, $y>0$, $\xk>0$, $\yk>0$  \cite{Tuybook}, which means that
\ifCLASSOPTIONpeerreview
\begin{align}\label{eq:ln_ineq}
\ln(x^{-1}+y^{-1}) \geq
\ln\left( \frac{1}{\xk} + \frac{1}{\yk}  \right) + 1  - \frac{1}{\xk+\yk} \left( \frac{\yk}{\xk} x + \frac{\xk}{\yk}y \right)
\end{align}
\else
\begin{align}\label{eq:ln_ineq}
\ln\left(\frac{1}{x} + \frac{1}{y} \right) &\geq \ln\left( \frac{1}{\xk} + \frac{1}{\yk}  \right) + 1   \notag\\
& - \frac{1}{\xk+\yk} \left( \frac{\yk}{\xk} x + \frac{\xk}{\yk}y \right)
\end{align}
\fi
Substituting $x = \|\bx\|^2$, $y = \|\bfy\|^2 + \sigma^2 $, $\xk = \|\bxk\|^2$, and $\yk = \|\byk\|^2 + \sigma^2 $, we have
\ifCLASSOPTIONpeerreview
\begin{eqnarray}\label{ap1}
\ln  \Big((\|\bx\|^2)^{-1}+(\|\bfy\|^2+\sigma^2)^{-1} \Big)&\geq &\ds\ln  \Big( (\|\bxk\|^2)^{-1}+(\|\byk\|^2+\sigma^2)^{-1} \Big)+1\nonumber\\
&&-\ds\frac{1}{\|\bxk\|^2+\|\byk\|^2+\sigma^2} \bigg(\frac{\|\byk\|^2+\sigma^2}{\|\bxk\|^2}\|\bx\|^2\nonumber\\
&&+\ds\frac{\|\bxk\|^2}{\|\byk\|^2+\sigma^2}(\|\bfy\|^2+\sigma^2)\bigg)
\end{eqnarray}
\else
\begin{eqnarray}\label{ap1}
\ln  \Big((\|\bx\|^2)^{-1}+(\|\bfy\|^2+\sigma^2)^{-1} \Big)&  \geq   &\notag \\
\ds\ln  \Big( (\|\bxk\|^2)^{-1}+(\|\byk\|^2+\sigma^2)^{-1} \Big)+1&&\nonumber\\
-\ds\frac{1}{\|\bxk\|^2+\|\byk\|^2+\sigma^2} \bigg(\frac{\|\byk\|^2+\sigma^2}{\|\bxk\|^2}\|\bx\|^2&&\nonumber\\
+\ds\frac{\|\bxk\|^2}{\|\byk\|^2+\sigma^2}(\|\bfy\|^2+\sigma^2)\bigg).&&
\end{eqnarray}
\fi
Next, as the functions $\ln(1/x)$ and $\|\bx\|^2$  are convex, it is true that
\begin{align}
\ln\left(\frac{1}{x} \right) &\ge \ln \left(\frac{1}{\xk} \right) -  \frac{x - \xk}{\xk} \label{ap12a} \\
\|\bx\|^2 &\ge 2 \Re\{(\bxk)^H\bx\}-\|\bxk\|^2.
\end{align}
By substituting $\|\bx\|^2$ and $\|\bxk\|^2$ in place of $\frac{1}{x}$ and $\frac{1}{\xk}$ in \eqref{ap12a}, we have the following inequality:
\begin{equation}\label{ap2}
\ln(\|\bx\|^2)\geq \ln(\|\bxk\|^2)+1-\frac{\|\bxk\|^2}{2 \Re\{(\bxk)^H\bx\}-\|\bxk\|^2}
\end{equation}
over the trust region (\ref{tr0}). Combining (\ref{ap1}) and (\ref{ap2}) leads to (\ref{es1})-(\ref{es1a}).
\section*{Appendix B: Proof for (\ref{newin})}
As function $g(x,t)\triangleq x^2/t$ is convex in $x>0$ and $t>0$, it is true that \cite{Tuybook}
$\frac{x^2}{t} \geq g(\bar{x},\bar{t})+\la \nabla g(\bar{x},\bar{t}), (x,t)-(\bar{x},\bar{t})\ra$
$=2\frac{\bar{x}}{\bar{t}}x-\frac{\bar{x}^2}{\bar{t}^2}t$.
Inequality (\ref{newin}) then follows by resetting $x\rightarrow \sqrt{x}$ and $\bar{x}\rightarrow \sqrt{\bar{x}}$.


\balance


\end{document}